\documentclass{article}
\usepackage{amsfonts}

\usepackage{amsmath}
\usepackage[dvips]{graphicx}

\newtheorem{theorem}{Theorem}

\newtheorem{lemma}[theorem]{Lemma}

\newtheorem{proposition}[theorem]{Proposition}

\input{tcilatex}

\begin{document}

\begin{titlepage}

\begin{center}
{\Large \textbf{\ The modular geometry }}

{\Large \textbf{\ of }}

{\Large \textbf{Random Regge Triangulations}}

\bigskip

{\large \textsl{M. Carfora}$\,^{b,}$\footnote{{\large 
mauro.carfora@pv.infn.it}},} \vspace{24pt}

{\large \textsl{C. Dappiaggi}$\,^{b,}$\footnote{{\large 
claudio.dappiaggi@pv.infn.it}},} \vspace{24pt}

\textsl{\large \textsl{A. Marzuoli}$\,^{b,}$\footnote{{\large 
annalisa.marzuoli@pv.infn.it}},} \vspace{24pt}

$^{b}$~Dipartimento di Fisica Nucleare e Teorica,

Universit\`{a} degli Studi di Pavia, \\[0pt]
via A. Bassi 6, I-27100 Pavia, Italy, \\[0pt]
and\\[0pt]
Istituto Nazionale di Fisica Nucleare, Sezione di Pavia, \\[0pt]
via A. Bassi 6, I-27100 Pavia, Italy
\end{center}

\begin{abstract}

We show that the introduction of triangulations with variable connectivity
and fluctuating egde-lengths (Random Regge Triangulations) allows for a
relatively simple and direct analyisis of the modular properties of 2
dimensional simplicial quantum gravity. In particular, we discuss in detail
an explicit bijection between the space of possible random Regge
triangulations (of given genus $g$ and with $N_{0}$ vertices) and a suitable
decorated version of the (compactified) moduli space of genus $g$ Riemann
surfaces with $N_{0}$ punctures  $\overline{\mathfrak{M}}_{g},_{N_{0}}$. Such an
analysis allows us to associate a Weil-Petersson metric with the set of 
random Regge triangulations and prove that the corresponding volume provides
the  dynamical triangulation partition function for pure gravity.
\end{abstract}

\bigskip


\bigskip


\end{titlepage}

\section{Introduction}

In this paper we discuss some aspects of 2-dimensional simplicial quantum
gravity by using surfaces endowed with triangulations with variable
connectivity and fluctuating egde-lengths. Such piecewise-linear (PL)
surfaces are not proper Regge triangulations, since their adjacency matrix
is not a priori fixed, nor they are dynamical triangulations, since they are
generated by glueing triangles which are not, in general, equilateral.
Lacking better names, we call them Random Regge Triangulations. They inherit
most of the nice features of their parent triangulations and put in a more
proper perspective some of the pathologies of Regge triangulations which
have often been misjudged. For instance, we show that most of these
pathologies have a modular origin and are naturally related to the
Weil-Petersson geometry of the (compactified) moduli space of genus $g$
Riemann surfaces with $N_{0}$ punctures $\overline{\mathfrak{M}}_{g},_{N_{0}}$,
(the number of punctures $N_{0}$ is the number of vertices of the
triangulations). As a matter of fact, the main feature of random Regge
triangulations is their allowing for a relatively simple and direct
analysis of the modular properties of 2 dimensional simplicial quantum
gravity. Usually, any such analysis requires techniques from the theory of
integrable system, matrix models, and conformal field theory. This
interdisciplinariety is one of the reason that motivates the success of
two-dimensional quantum gravity, however it tends to relegate simplicial
quantum gravity to the ancillary role of a regularization scheme. In such a
sense, it is important to establish a connection between moduli space theory
and simplicial quantum gravity which is more directly a consequence of the
use of simplicial methods, thus making their role more foundational. In
order to achieve such an objective, we explicitly map random Regge
triangulations into the theory of singular Euclidean structures. These
latter are represented by closed surfaces endowed with a metric which is
locally isometric to the euclidean plane except for a finite number of cone
singularities. Such a mapping allows for a rather straightforward bijection
between the space of possible random Regge triangulations (of given genus $g$
and with $N_{0}$ vertices) and a suitable decorated version of the moduli
space $\overline{\mathfrak{M}}_{g},_{N_{0}}$. The construction we present here
is partly related to the geometrical setup of the Witten-Kontsevich model
and the combinatorial parametrization of $\overline{\mathfrak{M}}_{g},_{N_{0}}$
in terms of ribbon graphs theory. However, even if strongly based on ribbon
graph theory, our approach requires a careful treatment of the role of the
deficit angles and the development of a technique for relating the Regge
geometry of the triangulations to the modular parameters of $\overline{\mathfrak{
M}}_{g},_{N_{0}}$. The main result of our analysis is the explicit
association of a Weil-Petersson metric to a Regge triangulation. With such a
metric at our disposal, we can formally evaluate the Weil-Petersson volume
over the space of all random Regge triangulations. By exploiting a recent
result of Manin and Zograf \cite{manin}, it is then an easy matter to show that such a
volume provides the (large $N_{0}$ asymptotics of the) dynamical
triangulation partition function for pure gravity.

Let us briefly summarize the content of the paper. 
In Section 2, after providing a few basic definitions, we describe the interplay between (random) Regge triangulations 
and the theory of singular Euclidean structures. To some extent this is 
the underlying rationale of the paper, allowing us to develop a dictionary between piecewise-linear surface geometry and standard complex 
analytical surface theory. Subsections 2.1, and 2.2 elaborate on this and related matters, such as the elegant Troyanov formulation of 
the Gauss-Bonnet theorem for triangulated surfaces and its implications. In particular, in subsection 2.3 we discuss the analytical meaning 
of zero-deficit angle degenerations in Regge calculus. Moreover, in subsection 2.4 we explicitly relate the Regge deficit angles to suitable 
moduli parameters in the complex uniformizations of the stars of the vertices of the triangulation. This result features prominently in the 
main applications of the paper, since it motivates the analysis of the geometry of Regge triangulations from the point of view of moduli theory. 
Clearly such an analysis must come to terms with the well-known correspondence between ribbon graph theory and combinatorial 
cell decompositions of the moduli space $\overline{\mathfrak{M}}_{g},_{N_{0}}$ which is at the basis of the Witten-Kontsevich model. In 
such a connection we discuss, in section 3, how we can naturally associate a ribbon graph with the 1-skeleton of a(random) Regge polytope.
We also elaborate on the stratified orbifold structure of the space of random Regge polytopes. In subsection 3.2 we show 
that these orbifold strata are naturally parametrized by dynamical triangulations. It is worthwhile stressing that such results can be seen as 
a rather elementary consequence of the remarkable analysis of the space of ribbon graphs due to M. Mulase and M. Penkawa \cite{mulase}. 
Our debt to the paper \cite{mulase} extends further in section 4,
 where we specialize some of their results on ribbon graphs theory in order to provide an explicit correspondence between (random)Regge triangulations an $N_0$-pointed Riemann surfaces (proposition 3, subsection 4.1). 
We come to full circle with moduli space theory when, by exploiting such a correspondence, we discuss deformations of Regge triangulations (subsection4.2). 
In such a setting, by examining the analytical aspects of Regge deformations, one is naturally led to consider the Weil-Petersson measure on the space of Regge polytopes. 
Among other things, such a construction further elucidates the deep geometrical nature of the degenerations in Regge calculus.
Finally, in subsection 4.3 we explicitly show that the (asymptotic) evaluation of the Weil-Petersson volume over the space of random Regge polytopes is strictly related to the canonical partition function of dynamical triangulation theory.

\section{Triangulated surfaces and Polytopes}

Let $T$ denote a $2$-dimensional simplicial complex with underlying
polyhedron $|T|$ and $f$- vector $(N_{0}(T),N_{1}(T),N_{2}(T))$, where $N_{i}(T)\in \mathbb{N}$ 
is the number of $i$-dimensional sub-simplices 
$\sigma ^{i}$ of $T$. Given a simplex $\sigma $ we denote by $st(\sigma )$,
(the star of $\sigma $), the union of all simplices of which $\sigma $ is a
face, and by $lk(\sigma )$, (the link of $\sigma $), is the union of all
faces $\sigma ^{f}$ of the simplices in $st(\sigma )$ such that $\sigma
^{f}\cap \sigma =\emptyset $. A Regge triangulation of a $2$
-dimensional PL manifold $M$, (without boundary), is a homeomorphism 
$|T_{l}|\rightarrow {M}$ where each face of $T$ is geometrically
realized by a rectilinear simplex of variable edge-lengths $l(\sigma
^{1}(k))$ (figure 1). A dynamical triangulation $|T_{l=a}|\rightarrow {M}$ is a
particular case of a Regge PL-manifold realized by rectilinear and
equilateral simplices of edge-length $l(\sigma ^{1}(k))=a$.

The metric structure of a Regge triangulation is locally Euclidean everywhere except
at the vertices $\sigma ^{0}$, (the \textit{bones}), where the sum of the
dihedral angles, $\theta (\sigma ^{2})$, of the incident triangles $\sigma
^{2}$'s is in excess (negative curvature) or in defect (positive curvature)
with respect to the $2\pi $ flatness constraint. The corresponding deficit
angle $\varepsilon $ is defined by $\varepsilon =2\pi -\sum_{\sigma
^{2}}\theta (\sigma ^{2})$, where the summation is extended to all $2$
-dimensional simplices incident on the given bone $\sigma ^{0}$. If $K_{T}^{0}$ 
denotes the $(0)$-skeleton of $|T_{l}|\rightarrow {M}$, (\emph{i.e.}, the collection of vertices of the triangulation), 
then $M\backslash{K_{T}^{0}}$ is a flat Riemannian manifold, and any point in the interior of
an $r$- simplex $\sigma ^{r}$ has a neighborhood homeomorphic to $B^{r}\times {C}(lk(\sigma ^{r}))$, 
where $B^{r}$ denotes the ball in $\mathbb{R}^{n}$ and ${C}(lk(\sigma ^{r}))$ is the cone over the link 
$lk(\sigma ^{r})$, (the product $lk(\sigma ^{r})\times \lbrack 0,1]$ with 
$lk(\sigma ^{r})\times \{1\}$ identified to a point). In particular, let
us denote by $C|lk(\sigma ^{0}(k))|$ the cone over the link of the vertex 
$\sigma ^{0}(k)$. On any such a disk $C|lk(\sigma ^{0}(k))|$ we can introduce
a locally uniformizing complex coordinate $\zeta (k)\in \mathbb{C}$ in terms
of which we can explicitly write down a conformal conical metric locally
characterizing the singular structure of $|T_{l}|\rightarrow M$, \emph{viz.
}, 
\begin{equation}
e^{2u}\left| \zeta (k)-\zeta _{k}(\sigma ^{0}(k))\right| ^{-2\left( \frac{
\varepsilon (k)}{2\pi }\right) }\left| d\zeta (k)\right| ^{2},  \label{cmetr}
\end{equation}
where $\varepsilon (k)$ is the corresponding deficit angle, and
$u:B^{2}\rightarrow \mathbb{R}$ is a continuous function ($C^{2}$ on 
$B^{2}-\{\sigma ^{0}(k)\}$) such that, for $\zeta (k)\rightarrow \zeta
_{k}(\sigma ^{0}(k))$, we have $\left| \zeta (k)-\zeta _{k}(\sigma
^{0}(k))\right| \frac{\partial u}{\partial \zeta (k)}$, and $\left| \zeta
(k)-\zeta _{k}(\sigma ^{0}(k))\right| \frac{\partial u}{\partial \overline{
\zeta }(k)}$ both $\rightarrow 0$, \cite{troyanov}. Up to the presence of $e^{2u}$, we
immediately recognize in such an expression the metric $g_{\theta (k)}$ of a
Euclidean cone of total angle $\theta (k)=2\pi -\varepsilon (k)$. The factor 
$e^{2u}$ allows to move within the conformal class of all metrics possessing
the same singular structure of the triangulated surface $|T_{l}|\rightarrow M$. 
We can profitably shift between the PL and the function theoretic point
of view by exploiting standard techniques of complex analysis, and making
contact with moduli space theory.

\subsection{Curvature assignments and divisors.} In the case of
dynamical triangulations, the picture simplifies considerably since the
deficit angles are generated by the numbers $\#\{\sigma ^{2}(h)\bot \sigma
^{0}(i)\}$\ of triangles incident on the $N_{0}(T)$ vertices, the \textit{
curvature assignments}, $\{q(k)\}_{k=1}^{N_{0}(T)}\in \mathbb{N}^{N_{0}(T)}$,

\begin{equation}
q(i)=\frac{2\pi -\varepsilon (i)}{\arccos (1/2)}.  \label{curvat}
\end{equation}
For a regular triangulation we have $q(k)\geq 3$, and since each triangle
has three vertices $\sigma ^{0}$, the set of integers $\{q(k)
\}_{k=1}^{N_{0}(T)}$ is constrained by

\begin{equation}
\sum_{k}^{N_{0}}q(k)=3N_{2}=6\left[ 1-\frac{\chi (M)}{N_{0}(T)}\right]
N_{0}(T),  \label{vincolo}
\end{equation}
where $\chi (M)$ denotes the Euler-Poincar\'{e} characteristic of the
surface, and where $6\left[ 1-\frac{\chi (M)}{N_{0}(T)}\right] $, ($\simeq 6$
for $N_{0}(T)>>1$),\ is the average value of the curvature assignments 
$\{q(k)\}_{k=1}^{N_{0}}$. More generally we shall consider semi-simplicial
complexes for which the constraint $q(k)\geq 3$ is removed. Examples of such
configurations are afforded by triangulations with pockets, where two
triangles are incident on a vertex, or by triangulations where the star of a
vertex may contain just one triangle. We shall refer to such extended
configurations as generalized (Regge and dynamical) triangulations.

The singular structure of the metric defined by (\ref{cmetr}) can be
naturally summarized in a formal linear combination of the points $\{\sigma
^{0}(k)\}$ with coefficients given by the corresponding deficit angles
(normalized to $2\pi $), i.e. in the \emph{real divisor }\cite{troyanov} 
\begin{equation}
Div(T)\doteq \sum_{k=1}^{N_{0}(T)}\left( -\frac{\varepsilon (k)}{2\pi }
\right) \sigma ^{0}(k)=\sum_{k=1}^{N_{0}(T)}\left( \frac{\theta (k)}{2\pi }
-1\right) \sigma ^{0}(k)
\end{equation}
supported on the set of bones $\{\sigma ^{0}(i)\}_{i=1}^{N_{0}(T)}$. Note
that the degree of such a divisor, defined by 
\begin{equation}
\left| Div(T)\right| \doteq \sum_{k=1}^{N_{0}(T)}\left( \frac{\theta (k)}{2\pi }-1\right) =-\chi (M)  \label{rediv}
\end{equation}
is, for dynamical triangulations, a rewriting of the combinatorial
constraint (\ref{vincolo}). In such a sense, the pair $(|T_{l=a}|\rightarrow
M,Div(T))$, or shortly, $(T,Div(T))$, encodes the datum of the triangulation 
$|T_{l=a}|\rightarrow M$ and of a corresponding set of curvature assignments 
$\{q(k)\}$ on the vertices $\{\sigma ^{0}(i)\}_{i=1}^{N_{0}(T)}$. The real
divisor $\left| Div(T)\right| $ characterizes the Euler class of the pair 
$(T,Div(T))$ and yields for a corresponding Gauss-Bonnet formula.
Explicitly, the Euler number associated with $(T,Div(T))$ is defined, \cite{troyanov},
by

\begin{equation}
e(T,Div(T))\doteq \chi (M)+|Div(T)\mathbf{|.}  \label{euler}
\end{equation}
and the Gauss-Bonnet formula reads \cite{troyanov}:

\begin{lemma}
(\textbf{Gauss-Bonnet for triangulated surfaces}) Let $(T,Div(T))$ be a
triangulated surface with divisor 
\begin{equation}
Div(T)\doteq \sum_{k=1}^{N_{0}(T)}\left( \frac{\theta (k)}{2\pi }-1\right)
\sigma ^{0}(k),
\end{equation}
associated with the vertices $\{\sigma ^{0}(k)\}_{k=1}^{N_{0}(T)}$. Let $ds^{2}$ 
be the conformal metric (\ref{cmetr}) representing the divisor $Div(T)$ . Then 
\begin{equation}\label{Euclass}
\frac{1}{2\pi }\int_{M}KdA=e(T,Div(T)),  
\end{equation}
where $K$ and $dA$ respectively are the curvature and the area element
corresponding to the local metric $ds_{(k)}^{2}.$
\end{lemma}

Note that such a theorem holds for any singular Riemann surface $\Sigma $
described by a divisor $Div(\Sigma )$ and not just for triangulated surfaces
\cite{troyanov}. Since for a Regge (dynamical) triangulation, we have 
$e(T_{a},Div(T))=0$, the Gauss-Bonnet formula implies

\begin{equation}
\frac{1}{2\pi }\int_{M}KdA=0.  \label{GaussB}
\end{equation}
Thus, a triangulation $|T_{l}|\rightarrow M$ naturally carries a conformally
flat structure. Clearly this is a rather obvious result, (since the metric
in $M-\{\sigma ^{0}(i)\}_{i=1}^{N_{0}(T)}$ is flat). However, it admits a
not-trivial converse (recently proved by M. Troyanov, but, in a sense, going
back to E. Picard) \cite{troyanov}, \cite{picard}:

\begin{theorem}
(\textbf{Troyanov-Picard}) Let $(\left( M,\mathcal{C}_{sg}\right) ,Div)$ be
a singular Riemann surface with a divisor such that $e(M,Div)=0$. Then there
exists on $M$ a unique (up to homothety) conformally flat metric
representing the divisor $Div$.
\end{theorem} 

\subsection{Conical Regge polytopes.}
Let us consider the (first) barycentric subdivision of $T$. The closed stars, in such a subdivision,
of the vertices of the original triangulation $T_{l}$ form a collection of 
$2$-cells $\{\rho ^{2}(i)\}_{i=1}^{N_{0}(T)}$ characterizing the \emph{
conical} Regge polytope $|P_{T_{l}}|\rightarrow {M}$ (and its rigid
equilateral specialization $|P_{T_{a}}|\rightarrow {M}$) barycentrically
dual to $|T_{l}|\rightarrow {M}$. If $(\lambda (k),\chi (k))$ denote polar
coordinates (based at $\sigma ^{0}(k)$) of a point $p\in \rho ^{2}(k)$, then 
$\rho ^{2}(k)$ is geometrically realized as the space 
\begin{equation}
\left. \left\{ (\lambda (k),\chi (k))\ :\lambda (k)\geq 0;\chi (k)\in 
\mathbb{R}/(2\pi -\varepsilon (k))\mathbb{Z}\right\} \right/ (0,\chi
(k))\sim (0,\chi ^{\prime }(k))
\end{equation}
endowed with the metric 
\begin{equation}
d\lambda (k)^{2}+\lambda (k)^{2}d\chi (k)^{2}.
\end{equation}
In other words, here we are not considering a rectilinear presentation of
the dual cell complex $P$ (where the PL-polytope is realized by flat
polygonal $2$-cells $\{\rho ^{2}(i)\}_{i=1}^{N_{0}(T)}$) but rather a
geometrical presentation $|P_{T_{l}}|\rightarrow {M}$ of $P$ where the $2$
-cells $\{\rho ^{2}(i)\}_{i=1}^{N_{0}(T)}$ retain the conical geometry
induced on the barycentric subdivision by the original metric (\ref{cmetr})
structure of $|T_{l}|\rightarrow {M}$ (figure 3).

\subsection{Degenerations: hyperbolic cusps and cylindrical ends.}\label{degenerations}
It is important to stress that whereas a Regge triangulation characterizes
a unique (up to automorphisms) singular Euclidean structure, this latter
actually allows for a more general type of metric triangulation. The point
is that some of the vertices associated with a singular Euclidean structure
can be characterized by deficit angles $\varepsilon (k)\rightarrow 2\pi$\emph{i.e.}, 
$\sum_{\sigma ^{2}(k)}\theta (\sigma ^{2}(k))=0$.
Such a situation
corresponds to having the cone $C|lk(\sigma ^{0}(k))|$ over the link 
$lk(\sigma ^{0}(k))$ realized by a Euclidean cone of angle $0$. This is a
natural limiting case in a Regge triangulation, (think of a vertex where
many long and thin triangles are incident), and it is usually discarded as
an unwanted pathology. However, there is really nothing pathological about
that, since the corresponding $2$-cell $\rho ^{2}(k)\in
|P_{T_{l}}|\rightarrow {M}$ can be naturally endowed with the conformal
Euclidean structure obtained from (\ref{cmetr}) by setting $\frac{
\varepsilon (k)}{2\pi }=1$, \emph{i.e. } 
\begin{equation}
e^{2u}\left| \zeta (k)-\zeta _{k}(\sigma ^{0}(k))\right| ^{-2}\left| d\zeta
(k)\right| ^{2},
\end{equation}
which (up to the conformal factor $e^{2u}$) is the flat metric on the
half-infinite cylinder $\mathbb{S}^{1}\times \mathbb{R}^{+}$ (a cylindrical
end). 
Alternatively, one may consider $\rho ^{2}(k)$ endowed with the
geometry of a hyperbolic cusp (figure 4), \emph{i.e.}, that of a half-infinite
cylinder $\mathbb{S}^{1}\times \mathbb{R}^{+}$ equipped with the hyperbolic
metric $\lambda (k)^{-2}(d\lambda (k)^{2}+d\chi (k)^{2})$. The triangles
incident on $\sigma ^{0}(k)$ are then realized as hyperbolic triangles with
the vertex $\sigma ^{0}(k)$ located at $\lambda (k)=\infty $ and
corresponding angle $\theta _{k}=0$\cite{judge}. Since the Poincar\'{e} metric on the
punctured disk $\{\zeta (k)\in C|\;0<\left| \zeta (k)-\zeta _{k}(\sigma
^{0}(k))\right| <1\}$ is 
\begin{equation}
\left( \left| \zeta (k)-\zeta _{k}(\sigma ^{0}(k))\right| \ln \frac{1}{
\left| \zeta (k)-\zeta _{k}(\sigma ^{0}(k))\right| }\right) ^{-2}\left|
d\zeta (k)\right| ^{2},
\end{equation}
one can shift from the Euclidean to the hyperbolic metric by setting 
\begin{equation}
e^{2u}=\left( \ln \frac{1}{\left| \zeta (k)-\zeta _{k}(\sigma
^{0}(k))\right| }\right) ^{-2},
\end{equation}
and the two points of view are strictly related. 

At any rate the presence of
hyperbolic cusps or cylindrical ends is consistent with a singular
Euclidean structure as long as the associated divisor satisfies the
topological constraint (\ref{rediv}),\emph{\ }which we can rewrite as 
\begin{equation}
\sum_{\{\frac{\varepsilon (k)}{2\pi }\neq 1\}}\left( -\frac{\varepsilon (k)}{
2\pi }\right) =2g-2+\#\left\{ \sigma ^{0}(h)|\;\frac{\varepsilon (h)}{2\pi }
=1\right\} .
\end{equation}
In particular, we can have the limiting case of the singular Euclidean
structure associated with a genus $g$ surface triangulated with $N_{0}-1$
hyperbolic vertices $\{\sigma ^{0}(k)\}_{k=1}^{N_{0}-1}$ (or, equivalently,
with $N_{0}-1$ cylindrical ends) and just one standard conical vertex, $
\sigma ^{0}(N_{0})$, supporting the deficit angle 
\begin{equation}\label{lastmarked}
-\frac{\varepsilon (N_{0})}{2\pi }=2g-2+(N_{0}-1).  
\end{equation}

\subsection{Conical sectors and moduli.}\label{conical sectors}
We can represent finer details of the geometry of $(U_{\rho ^{2}(k)},ds_{(k)}^{2})$ by opening the
cone into its constituent conical sectors. To motivate such a
representation, let $W_{\alpha }(k)$, $\alpha =1,...,q(k)$, be the
barycenters of the edges $\sigma ^{1}(\alpha )\in |T_{l}|\rightarrow {M}$
incident on $\sigma ^{0}(k)$, and intercepting the boundary $\partial
(\rho ^{2}(k))$ of the polygonal cell $\rho ^{2}(k)$. Denote by $l(\partial
(\rho ^{2}(k)))$ the length of $\partial (\rho ^{2}(k))$, and by $\widehat{L}
_{\alpha }(k)$ the length of the polygonal $\partial (\rho ^{2}(k))$
between the points $W_{\alpha }(k)$ and $W_{\alpha +1}(k)$ (with $\alpha $
defined $\func{mod}q(k)$). In the uniformization $\zeta (k)$ of $
C|lk(\sigma ^{0}(k))|$, the points $\{W_{\alpha }(k)\}$ characterize a
corresponding set of points on the circumference $\{\zeta (k)\in \mathbb{C}
\;|\;\left| \zeta (k)\right| =l(\partial (\rho ^{2}(k)))\}$, (for
simplicity, we have set $\zeta _{k}(\sigma ^{0}(k))=0$), and an associated
set of $q(k)$ generators $\{\overline{W_{\alpha }(k)\sigma ^{0}(k)}\}$ 
on the cone $(U_{\rho ^{2}(k)},ds_{(k)}^{2})$. Such generators mark $q(k)$ conical
sectors 
\begin{equation}
S_{\alpha }(k)\doteq \left( c_{\alpha }(k),\frac{l(\partial (\rho ^{2}(k)))}{
\theta (k)},\vartheta _{\alpha }(k)\right) ,
\end{equation}
with base 
\begin{equation}
c_{\alpha }(k)\doteq \left\{ \left| \zeta (k)\right| =l(\partial (\rho
^{2}(k))),\arg W_{\alpha }(k)\leq \arg \zeta (k)\leq \arg W_{\alpha
+1}(k)\right\} ,  \label{consect}
\end{equation}
slant radius $\frac{l(\partial (\rho ^{2}(k)))}{\theta (k)}$, and with
angular opening 
\begin{equation}
\vartheta _{\alpha }(k)\doteq \frac{\widehat{L}_{\alpha }(k)}{l(\partial
(\rho ^{2}(k)))}\theta (k),  \label{angles}
\end{equation}
where $\theta (k)=2\pi -\varepsilon (k)$ is the given conical angle. Since $
\sum_{\alpha =1}^{q(k)}\vartheta _{\alpha }(k)=\theta (k)$, the $\{\vartheta
_{\alpha }(k)\}$ are the representatives, in the uniformization $(U_{\rho
^{2}(k)},ds_{(k)}^{2})$, of the $q(k)$ vertex angles generating the deficit
angle $\varepsilon (k)$ of $C|lk(\sigma ^{0}(k))|$. In particular, we can
formally represent the cone $(U_{\rho ^{2}(k)},ds_{(k)}^{2})$ as 
\begin{equation}
(U_{\rho ^{2}(k)},ds_{(k)}^{2})=\cup _{\alpha =1}^{q(k)}S_{\alpha }(k).
\end{equation}
If we split open the vertex of the cone and of the associated conical
sectors $S_{\alpha }(k)$, then the conical geometry of $(U_{\rho
^{2}(k)},ds_{(k)}^{2})$ can be equivalently described by a cylindrical strip
of height $\frac{l(\partial (\rho ^{2}(k)))}{\theta (k)}$ decorating the
boundary of $\rho ^{2}(k)$. Each sector $S_{\alpha }(k)$ in the cone gives
rise, in such a picture, to a rectangular region in the cylindrical strip (figure 5).
It is profitable to explicitly represent any such a region, in the complex
plane of the variable $z=x+\sqrt{-1}y$, upside down according to 
\begin{equation}
R_{\vartheta _{\alpha }(k)}(k)\doteq \left\{ z\in \mathbb{C}|\;0\leq x\leq 
\frac{l(\partial (\rho ^{2}(k)))}{\theta (k)},0\leq y\leq \widehat{L}
_{\alpha }(k)\right\} .
\end{equation}
We can go a step further, and by means of the conformal transformation 
\begin{equation}
W_{\alpha }(k)=\exp \left[ \frac{2\pi \sqrt{-1}\theta (k)}{l(\partial (\rho
^{2}(k)))}z\right] ,
\end{equation}
we can map the rectangle $R_{\vartheta _{\alpha }(k)}(k)$ into the annulus (figure 6)
\begin{equation}
\Delta _{\vartheta _{\alpha }(k)}(k)\doteq \{W(k)\in \mathbb{C}|\;|t_{\alpha
}(k)|<\left| W(k)\right| <1\},  \label{annulus}
\end{equation}
where 
\begin{equation}
|t_{\alpha }(k)|\doteq \exp \left[ -2\pi \vartheta _{\alpha }(k)\right]. 
\end{equation}

\noindent Note that 
\begin{equation}\label{modulann}
\frac{1}{2\pi }\ln \left( \frac{1}{|t_{\alpha }(k)|}\right) =\vartheta
_{\alpha }(k), 
\end{equation}
is the modulus of $\Delta _{\vartheta _{\alpha }(k)}(k)$.

Such a remark motivates the analysis of the geometry of (random) Regge triangulations
from the point of view of moduli theory. In this connection note that for a
given set of perimeters $\{l(\partial (\rho ^{2}(k)))\}_{k=1}^{N_{0}}$ and
deficit angles $\{\varepsilon (k)\}_{k=1}^{N_{0}}$ there are $
(N_{1}(T)-N_{0}(T))$ free angles $\vartheta _{\alpha }(k)$. This follows by
observing that, for given $l(\partial (\rho ^{2}(k)))$ and $\varepsilon (k)$
, the angles $\vartheta _{\alpha }(k)$ are characterized (see (\ref{angles}
)) by the $\widehat{L}_{\alpha }(k)$. These latter are in a natural
correspondence with the $N_{1}$ edges of  $|P_{T_{l}}|\rightarrow {M}$,
and among them we have the $N_{0}$ constraints $\sum_{\alpha }^{q(k)}\widehat{
L}_{\alpha }(k)=l(\partial (\rho ^{2}(k)))$. From $
N_{0}(T)-N_{1}(T)+N_{2}(T)=2-2g$, and the relation $2N_{1}(T)=3N_{2}(T)$
associated with the trivalency, we get $N_{1}(T)-N_{0}(T)=2N_{0}(T)+6g-6$,
which exactly corresponds to the real dimension of the moduli space $\mathfrak{M}
_{g},_{N_{0}}$ of genus $g$ Riemann surfaces $((M;N_{0}),\mathcal{C})$\ with 
$N_{0}$ punctures, (see below). 

\section{\noindent \textbf{Ribbon graphs on Regge Polytopes}}

The geometrical realization of the $1$-skeleton of the conical Regge
polytope $|P_{T_{l}}|\rightarrow {M}$ is a $3$-valent graph 
\begin{equation}
\Gamma =(\{\rho ^{0}(k)\},\{\rho ^{1}(j)\})
\end{equation}
where the vertex set $\{\rho ^{0}(k)\}_{k=1}^{N_{2}(T)}$ is identified with
the barycenters of the triangles $\{\sigma ^{o}(k)\}_{k=1}^{N_{2}(T)}\in
|T_{l}|\rightarrow M$, whereas each edge $\rho ^{1}(j)\in \{\rho
^{1}(j)\}_{j=1}^{N_{1}(T)}$ is generated by two half-edges $\rho ^{1}(j)^{+}$
and $\rho ^{1}(j)^{-}$ joined through the barycenters $\{W(h)
\}_{h=1}^{N_{1}(T)}$ of the edges $\{\sigma ^{1}(h)\}$ belonging to the
original triangulation $|T_{l}|\rightarrow M$. If we formally introduce a
ghost-vertex of a degree $2$ at each middle point $\{W(h)\}_{h=1}^{N_{1}(T)}$
, then the actual graph naturally associated to the $1$-skeleton of 
$|P_{T_{l}}|\rightarrow {M}$ is the edge-refinement \cite{mulase} of $\Gamma =(\{\rho^{0}(k)\},\{\rho ^{1}(j)\})$, \emph{i.e.} 
\begin{equation}
\Gamma _{ref}=\left( \{\rho^{0}(k)\}\bigsqcup_{h=1}^{N_{1}(T)}\{W(h)\},\{\rho^{1}(j)^{+}\}\bigsqcup_{j=1}^{N_{1}(T)}\{\rho ^{1}(j)^{-}\}\right),
\end{equation}
as it can be seen in figure 7.

\noindent The natural automorphism group $Aut(P_{l})$ of \ $|P_{T_{l}}|\rightarrow {M}$,
(\emph{i.e.}, the set of bijective maps $\Gamma =(\{\rho ^{0}(k)\},\{\rho
^{1}(j)\})\rightarrow \widetilde{\Gamma }=(\widetilde{\{\rho ^{0}(k)\}},
\widetilde{\{\rho ^{1}(j)\}}$ preserving the incidence relations defining
the graph structure), is the automorphism group of its edge refinement \cite{mulase}, 
$Aut(P_{l})\doteq Aut(\Gamma _{ref})$. The locally uniformizing complex
coordinate $\zeta (k)\in \mathbb{C}$ in terms of which we can explicitly
write down the singular Euclidean metric (\ref{cmetr}) around each vertex $
\sigma ^{0}(k)\in $ $|T_{l}|\rightarrow M$, provides a (counterclockwise)
orientation in the $2$-cells of $|P_{T_{l}}|\rightarrow {M}$. Such an
orientation gives rise to a cyclic ordering on the set of half-edges $\{\rho
^{1}(j)^{\pm }\}_{j=1}^{N_{1}(T)}$ incident on the vertices $\{\rho
^{0}(k)\}_{k=1}^{N_{2}(T)}$. According to these remarks, the $1$-skeleton
of $|P_{T_{l}}|\rightarrow {M}$ is a ribbon (or fat) graph \cite{ambjorn}, 
a graph $\Gamma $ together with a cyclic ordering on the set of half-edges
incident to each vertex of $\Gamma$. Conversely, any ribbon graph $\Gamma 
$ characterizes an oriented surface $M(\Gamma )$ with boundary possessing $
\Gamma $ as a spine, (\emph{i.e.}, the inclusion $\Gamma \hookrightarrow
M(\Gamma )$ is a homotopy equivalence). In this way (the edge-refinement of)
the $1$-skeleton of a generalized conical Regge polytope $
|P_{T_{l}}|\rightarrow {M}$ is in a one-to-one correspondence with trivalent
metric ribbon graphs. The set of all such trivalent ribbon graph $\Gamma $
with given edge-set $e(\Gamma )$ can be characterized \cite{mulase}, \cite{looijenga} as a space
homeomorphic to $\mathbb{R}_{+}^{|e(\Gamma )|}$, ($|e(\Gamma )|$ denoting
the number of edges in $e(\Gamma )$), topologized by the standard $
\epsilon $-neighborhoods $U_{\epsilon }\subset $ $\mathbb{R}
_{+}^{|e(\Gamma )|}$. The automorphism group $Aut(\Gamma )$ acts naturally
on such a space via the homomorphism $Aut(\Gamma )\rightarrow \mathfrak{G}
_{e(\Gamma )}$, where $\mathfrak{G}_{e(\Gamma )}$ denotes the symmetric group
over $|e(\Gamma )|$ elements, and the resulting quotient space $\mathbb{R}
_{+}^{|e(\Gamma )|}/Aut(\Gamma )$ is a differentiable orbifold.

\subsection{The space of 1-skeletons of Regge polytopes.}
Let $Aut_{\partial }(P_{l})\subset Aut(P_{l})$, denote the subgroup of ribbon
graph automorphisms of the (trivalent) $1$-skeleton $\Gamma $ of $
|P_{T_{l}}|\rightarrow {M}$ that preserve the (labeling of the) boundary
components of $\Gamma $. Then, the space $K_{1}RP_{g,N_{0}}^{met}$ of $1$
-skeletons of conical Regge polytopes $|P_{T_{l}}|\rightarrow {M}$, with $
N_{0}(T)$ labelled boundary components, on a surface $M$ of genus $g$ can be
defined by \cite{mulase}
\begin{equation}
K_{1}RP_{g,N_{0}}^{met}=\bigsqcup_{\Gamma \in RGB_{g,N_{0}}}\frac{\mathbb{R}
_{+}^{|e(\Gamma )|}}{Aut_{\partial }(P_{l})},  \label{DTorb}
\end{equation}
where the disjoint union is over the subset of all trivalent ribbon graphs
(with labelled boundaries) satisfying the topological stability condition $
2-2g-N_{0}(T)<0$, and which are dual to generalized triangulations. It
follows, (see \cite{mulase} theorems 3.3, 3.4, and 3.5), that the set $
K_{1}RP_{g,N_{0}}^{met}$ is locally modelled on a stratified space
constructed from the components (rational orbicells) $\mathbb{R}
_{+}^{|e(\Gamma )|}/Aut_{\partial }(P_{l})$ by means of a (Whitehead)
expansion and collapse procedure for ribbon graphs, which amounts to
collapsing edges and coalescing vertices, (the Whitehead move in $
|P_{T_{l}}|\rightarrow {M}$ is the dual of the familiar flip move \cite{ambjorn} for
triangulations). Explicitly, if $l(t)=tl$ is the length of an edge $\rho
^{1}(j)$ of a ribbon graph $\Gamma _{l(t)}\in $ $K_{1}RP_{g,N_{0}}^{met}$,
then, as $t\rightarrow 0$, we get the metric ribbon graph $\widehat{\Gamma }$
which is obtained from $\Gamma _{l(t)}$ by collapsing the edge $\rho ^{1}(j)$
. By exploiting such construction, we can extend the space $
K_{1}RP_{g,N_{0}}^{met}$ to a suitable closure $\overline{K_{1}RP}
_{g,N_{0}}^{met}$ \cite{looijenga}, (this natural topology on $K_{1}RP_{g,N_{0}}^{met}$
shows that, at least in two-dimensional quantum gravity, the set of Regge
triangulations with \emph{fixed connectivity} does not explore the full
configurational space of the theory). The open cells of 
$K_{1}RP_{g,N_{0}}^{met}$, being associated with trivalent graphs, have
dimension provided by the number $N_{1}(T)$ of edges of $|P_{T_{l}}|
\rightarrow {M}$, \emph{i.e.} 
\begin{equation}
\dim \left[ K_{1}RP_{g,N_{0}}^{met}\right] =N_{1}(T)=3N_{0}(T)+6g-6.
\end{equation}
There is a natural projection 
\begin{gather}
p:K_{1}RP_{g,N_{0}}^{met}\longrightarrow \mathbb{R}_{+}^{N_{0}(T)} \\
\Gamma \longmapsto p(\Gamma )=(l_{1},...,l_{N_{0}(T)}),  \notag
\end{gather}
where $(l_{1},...,l_{N_{0}(T)})$ denote the perimeters of the polygonal
2-cells $\{\rho ^{2}(j)\}$ of $|P_{T_{l}}|\rightarrow {M}$. With respect to
the topology on the space of metric ribbon graphs, the orbifold 
$K_{1}RP_{g,N_{0}}^{met}$ endowed with such a projection acquires the
structure of a cellular bundle. For a given sequence $\{l(\partial (\rho
^{2}(k)))\}$, the fiber 
\begin{equation}
p^{-1}(\{l(\partial (\rho ^{2}(k)))\})=\left\{ |P_{T_{l}}|\rightarrow {M}\in
K_{1}RP_{g,N_{0}}^{met}:\{l_{k}\}=\{l(\partial (\rho ^{2}(k)))\}\right\}
\end{equation}
is the set of all generalized conical Regge polytopes with the given set of
perimeters. If we take into account the $N_{0}(T)$ constraints associated
with the perimeters assignments, it follows that the fibers $p^{-1}(\{l(\partial (\rho ^{2}(k)))\})$ have dimension provided by 
\begin{equation}
\dim \left[ p^{-1}(\{l(\partial (\rho ^{2}(k)))\}\right] =2N_{0}(T)+6g-6,
\end{equation}
which again corresponds to the real dimension of the moduli space $\mathfrak{M}_{g},_{N_{0}}$ of $N_{0}$-pointed Riemann surfaces of genus $g$.

\subsection{Orbifold labelling and dynamical triangulations.}
Let us denote by 
\begin{equation}
\Omega _{T_{a}}\doteq \frac{\mathbb{R}_{+}^{|e(\Gamma )|}}{Aut_{\partial
}(P_{T_{a}})}  \label{omega}
\end{equation}
the rational cell associated with the 1-skeleton of the conical polytope $
|P_{T_{a}}|\rightarrow {M}$ dual to a dynamical triangulation $
|T_{l=a}|\rightarrow M$. The orbicell (\ref{omega}) contains the ribbon
graph associated with $|P_{T_{a}}|\rightarrow {M}$ and all (trivalent)
metric ribbon graphs $|P_{T_{L}}|\rightarrow {M}$ with the same
combinatorial structure of $|P_{T_{a}}|\rightarrow {M}$ but with all
possible length assignments $\{l(\rho ^{1}(h))\}_{h=1}^{N_{1}(T)}$ associated
with the corresponding set of edges $\{\rho ^{1}(h)\}_{1}^{N_{1}(T)}$. The
orbicell $\Omega _{T_{a}}$ is naturally identified with the convex
polytope (of dimension $(2N_{0}(T)+6g-6)$) in $\mathbb{R}_{+}^{N_{1}(T)}$
defined by 
\begin{equation}
\left\{ \{l(\rho ^{1}(j))\}\in \mathbb{R}_{+}^{N_{1}(T)}:
\sum_{j=1}^{q(k)}A_{(k)}^{j}(T_{a})l(\rho ^{1}(j))\,=\frac{\sqrt{3}}{3}
aq(k),\;k=1,...,N_{0}\;\right\} ,  \label{strata}
\end{equation}
where $A_{(k)}^{j}(T_{a})$ is a $(0,1)$ indicator matrix, depending on the
given dynamical triangulation $|T_{l=a}|\rightarrow M$, with $
A_{(k)}^{j}(T_{a})=1$ if the edge $\rho ^{1}(j)$ belongs to $\partial (\rho
^{2}(k))$, and $0$ otherwise, and $\frac{\sqrt{3}}{3}aq(k)$ is the perimeter
length $l(\partial (\rho ^{2}(k)))$ in terms of the corresponding curvature
assignment $q(k)$. Note that $|P_{T_{a}}|\rightarrow {M}$ appears as the
barycenter of such a polytope.

Since the cell decomposition (\ref{DTorb}) of the space of trivalent metric
ribbon graphs $K_{1}RP_{g,N_{0}}^{met}$ depends only on the combinatorial
type of the ribbon graph, we can use the equilateral polytopes $
|P_{T_{a}}|\rightarrow {M}$, dual to dynamical triangulations, as the set
over which the disjoint union in (\ref{DTorb}) runs. Thus we can write 
\begin{equation}
K_{1}RP_{g,N_{0}}^{met}=\bigsqcup_{\mathcal{DT}(N_{0})}\Omega _{T_{a}},
\end{equation}
where 
\begin{equation}
\mathcal{DT}_{g}\left( N_{0}\right) \doteq \left\{ |T_{l=a}|\rightarrow
M\;:(\sigma ^{0}(k))\;k=1,...,N_{0}(T)\right\}
\end{equation}
denote the set of distinct generalized dynamically triangulated surfaces
of genus $g$, with a given set of $N_{0}(T)$ ordered labelled vertices.  

Note that, even if the set $\mathcal{DT}_{g}\left( N_{0}\right) $ can be
considered (through barycentrical dualization) a well-defined subset of $
K_{1}RP_{g,N_{0}}^{met}$, it is not an orbifold over $\mathbb{N}$ \cite{mulase2}. For this
latter reason, the analysis of the metric stuctures over (generalized)
polytopes requires the use of the full orbicells $\Omega _{T_{a}}$ and we
cannot limit our discussion to equilateral polytopes.

\subsection{The ribbon graph parametrization of the
moduli space} 
We start by recalling that the moduli space $\mathfrak{M}
_{g},_{N_{0}}$ of genus $g$ Riemann surfaces with $N_{0}$ punctures is a
dense open subset of a natural compactification (Knudsen-Deligne-Mumford )
in a connected, compact complex orbifold denoted by $\overline{\mathfrak{M}}
_{g},_{N_{0}}$. This latter is, by definition, the moduli space of stable $
N_{0}$-pointed curves of genus $g$, where a stable curve is a compact
Riemann surface with at most ordinary double points such that all of its parts are
hyperbolic. The closure $\partial \mathfrak{M}_{g},_{N_{0}}$ of $\mathfrak{M}
_{g},_{N_{0}}$ in $\overline{\mathfrak{M}}_{g},_{N_{0}}$ consists of stable
curves with double points, and gives rise to a stratification decomposing $
\overline{\mathfrak{M}}_{g},_{N_{0}}$ into subvarieties. By definition, a
stratum of codimension $k$ is the component of $\overline{\mathfrak{M}}
_{g},_{N_{0}}$ parametrizing stable curves (of fixed topological type)
with $k$ double points.

The complex analytic geometry of the space of conical Regge polytopes which
we will discuss in the next section generalizes the well-known bijection (a
homeomorphism of orbifolds) between the space of metric ribbon graphs $
K_{1}RP_{g,N_{0}}^{met}$ (which forgets the conical geometry) and the moduli
space $\mathfrak{M}_{g},_{N_{0}}$ of genus $g$ Riemann surfaces $((M;N_{0}),
\mathcal{C})$ with $N_{0}(T)$ punctures \cite{mulase}, \cite{looijenga}. This bijection
results in a local parametrization of $\mathfrak{M}_{g},_{N_{0}}$ defined by 
\begin{gather}
h:K_{1}RP_{g,N_{0}}^{met}\rightarrow \mathfrak{M}_{g},_{N_{0}}\times {R}_{+}^{N}
\label{bijec} \\
\Gamma \longmapsto \lbrack ((M;N_{0}),\mathcal{C}),l_{i}]  \notag
\end{gather}
where $(l_{1},...,l_{N_{0}})$ is an ordered n-tuple of positive real numbers
and $\Gamma $ is a metric ribbon graphs with $N_{0}(T)$ labelled boundary
lengths $\{l_{i}\}$ (figure 8). 

If $\overline{K_{1}RP}_{g,N_{0}}^{met}$ is the closure
of $K_{1}RP_{g,N_{0}}^{met}$, then the bijection $h$ extends to $\overline{
K_{1}RP}_{g,N_{0}}^{met}\rightarrow \overline{\mathfrak{M}}_{g},_{N_{0}}\times {R
}_{+}^{N_{0}}$ in such a way that a ribbon graph $\Gamma \in \overline{RGP}
_{g,N_{0}}^{met}$ is mapped in two (stable) surfaces $M_{1}$ and $M_{2}$
with $N_{0}(T)$ punctures if and only if there exists an homeomorphism
between $M_{1}$ and $M_{2}$ preserving the (labelling of the) punctures, and
is holomorphic on each irreducible component containing one of the punctures.

According to Kontsevich \cite{kontsevich}, corresponding to each marked polygonal 2-cells $\{\rho ^{2}(k)\}$ of $|P_{T_{l}}|\rightarrow {M}$ 
there is a further (combinatorial) bundle map 
\begin{equation}
\mathcal{CL}_{k}\rightarrow K_{1}RP_{g,N_{0}}^{met}  \label{combundle}
\end{equation}
whose fiber over $(\Gamma ,\rho ^{2}(1),...,\rho ^{2}(N_{0}))$ is provided
by the boundary cycle $\partial \rho ^{2}(k)$, (recall that each boundary $\partial \rho ^{2}(k)$ 
comes with a positive orientation). To any such cycle
one associates \cite{looijenga}, \cite{kontsevich} the corresponding perimeter map $l(\partial (\rho^{2}(k)))=\sum $\ $l(\rho ^{1}(h_{\alpha }))$ 
which then appears as defining a natural connection on $\mathcal{CL}_{k}$. The piecewise smooth 2-form
defining the curvature of such a connection, 
\begin{equation}
\omega _{k}(\Gamma )=\sum_{1\leq h_{\alpha }<h_{\beta }\leq q(k)-1}d\left( 
\frac{l(\rho ^{1}(h_{\alpha }))}{l(\partial \rho ^{2}(k))}\right) \wedge
d\left( \frac{l(\rho ^{1}(h_{\beta }))}{l(\partial \rho ^{2}(k))}\right) ,
\label{chern}
\end{equation}
is invariant under rescaling and cyclic permutations of the $l(\rho
^{1}(h_{\mu }))$, and is a combinatorial representative of the Chern class
of the line bundle $\mathcal{CL}_{k}$.

\bigskip

It is important to stress that even if ribbon graphs can be thought of as
arising from Regge polytopes (with variable connectivity), the morphism (\ref
{bijec}) only involves the ribbon graph structure and the theory can be (and
actually is) developed with no reference at all to a particular underlying
triangulation. In such a connection, the role of dynamical triangulations
has been slightly overemphasized, they simply provide a convenient way of
labelling the different combinatorial strata of the mapping (\ref{bijec}),
but, by themselves they do not define a combinatorial parametrization of $
\overline{\mathfrak{M}}_{g},_{N_{0}}$ for any finite $N_{0}$. However, it is
very useful, at least for the purposes of quantum gravity, to remember the
possible genesis of a ribbon graph from an underlying triangulation and be
able to exploit the further information coming from the associated conical
geometry. Such an information cannot be recovered from the ribbon graph
itself (with the notable exception of equilateral ribbon graphs, which can
be associated with dynamical triangulations), and must be suitably codified
by adding to the boundary lengths $\{l_{i}\}$ of the graph a further
decoration. This can be easily done by explicitly connecting Regge polytopes
to punctured Riemann surfaces.

\section{\textbf{Regge polytopes and punctured Riemann surfaces. }}

As suggested by (\ref{cmetr}), the polyhedral metric associated with the
vertices $\{\sigma ^{0}(i)\}$ of a (generalized) Regge triangulation $
|T_{l}|\rightarrow M$, can be conveniently described in terms of complex
function theory. We can extend the ribbon graph
uniformization of \cite{mulase} and associate with $|P_{T_{l}}|\rightarrow {M}$ a
complex structure $((M;N_{0}),\mathcal{C})$ (a punctured Riemann surface)
which is, in a well-defined sense, dual to the structure (\ref{cmetr})
generated by $|T_{l}|\rightarrow M$. Let $\rho ^{2}(k)$ be the generic
two-cell $\in |P_{T_{l}}|\rightarrow {M}$ barycentrically dual to the vertex 
$\sigma ^{0}(k)\in |T_{l}|\rightarrow M$. To the generic edge $\rho
^{1}(h) $ of $\rho ^{2}(k)$ we associate a complex uniformizing
coordinate $z(h)$ defined in the strip 
\begin{equation}
U_{\rho ^{1}(h)}\doteq \{z(h)\in \mathbb{C}\;|\;0<\func{Re}z(h)<l(\rho
^{1}(h))\},
\end{equation}
$l(\rho ^{1}(h))$ being the length of the edge considered. The uniformizing
coordinate $w(j)$, corresponding to the generic $3$-valent vertex $\rho
^{0}(j)\in \rho ^{2}(k)$, is defined in the open set 
\begin{equation}
U_{\rho ^{0}(j)}\doteq \{w(j)\in \mathbb{C}\;|\;|w(j)|<\delta ,\;w(j)[\rho
^{0}(j)]=0\},
\end{equation}
where $\delta >0$ is a suitably small constant. Finally, the two-cell $\rho
^{2}(k)$ is uniformized in the unit disk 
\begin{equation}
U_{\rho ^{2}(k)}\doteq \{\zeta (k)\in \mathbb{C}\;|\;|\zeta (k)|<1,\;\zeta
(k)[\sigma ^{0}(k)]=0\},
\end{equation}
where $\sigma ^{0}(k)$ is the vertex $\in |T_{l}|\rightarrow M$ 
corresponding to the given two-cell (figure 9).

The various uniformizations $\{w(j),U_{\rho ^{0}(j)}\}_{j=1}^{N_{2}(T)}$, $
\{z(h),U_{\rho ^{1}(h)}\}_{h=1}^{N_{1}(T)}$, and $\{\zeta (k),U_{\rho
^{2}(k)}\}_{k=1}^{N_{0}(T)}$ can be coherently glued together by noting that
to each edge $\rho ^{1}(h)\in $ $\rho ^{2}(k)$ we can associate the
standard quadratic differential on $U_{\rho ^{1}(h)}$ given by 
\begin{equation}\label{foliat}
\phi (h)|_{\rho ^{1}(h)}=dz(h)\otimes dz(h).  
\end{equation}
Such $\phi (h)|_{\rho ^{1}(h)}$ can be extended to the remaining local
uniformizations $U_{\rho ^{0}(j)}$, and $U_{\rho ^{2}(k)}$, by exploiting
a classic result in Riemann surface theory according to which a quadratic
differential $\phi $ has a finite number of zeros $n_{zeros}(\phi )$ with
orders $k_{i}$ and a finite number of poles $n_{poles}(\phi )$ of order $
s_{i}$ such that 
\begin{equation}
\sum_{i=1}^{n_{zero}(\phi )}k_{i}-\sum_{i=1}^{n_{pole}(\phi )}s_{i}=4g-4.
\label{quadrel}
\end{equation}
In our case we must have $n_{zeros}(\phi )=N_{2}(T)$ with $k_{i}=1$,
(corresponding to the fact that the $1$-skeleton of $|P_{l}|\rightarrow M$
is a trivalent graph), and $n_{poles}(\phi )=$\ $N_{0}(T)$ with $
s_{i}=s\;\forall i$, for a suitable positive integer $s$. According to such
remarks (\ref{quadrel}) reduces to 
\begin{equation}\label{poles}
N_{2}(T)-sN_{0}(T)=4g-4. 
\end{equation}
>From the Euler relation $N_{0}(T)-N_{1}(T)+N_{2}(T)=2-2g$, and $
2N_{1}(T)=3N_{2}(T)$ we get $N_{2}(T)-2N_{0}(T)=4g-4$. This is consistent
with (\ref{poles}) if and only if $s=2$. Thus the extension $\phi $ of $
\phi (h)|_{\rho ^{1}(h)}$ along the $1$-skeleton of $|P_{l}|\rightarrow M$
must have $N_{2}(T)$ zeros of order $1$ corresponding to the trivalent
vertices $\{\rho ^{0}(j)\}$\ of $|P_{l}|\rightarrow M$ and $N_{0}(T)$
quadratic poles corresponding to the polygonal cells $\{\rho ^{2}(k)\}$ of
perimeter lengths $\{l(\partial (\rho ^{2}(k)))\}$. Around a zero of order
one and a pole of order two, every (Jenkins-Strebel \cite{strebel}) quadratic differential 
$\phi$ has a canonical local structure which (along with (\ref{foliat})) is
given by \cite{mulase}\cite{strebel} 
\begin{equation}
(|P_{T_{l}}|\rightarrow {M)\rightarrow }\phi \doteq \left\{ 
\begin{tabular}{l}
$\phi (h)|_{\rho ^{1}(h)}=dz(h)\otimes dz(h),$ \\ 
$\phi (j)|_{\rho ^{0}(j)}=\frac{9}{4}w(j)dw(j)\otimes dw(j),$ \\ 
$\phi (k)|_{\rho ^{2}(k)}=-\frac{\left[ l(\partial (\rho ^{2}(k)))\right]
^{2}}{4\pi ^{2}\zeta ^{2}(k)}d\zeta (k)\otimes d\zeta (k),$
\end{tabular}
\right.  \label{differ}
\end{equation}
where $\{\rho ^{0}(j),\rho ^{1}(h),\rho ^{2}(k)\}$ runs over the set of
vertices, edges, and $2$-cells of $|P_{T_{l}}|\rightarrow M$. Since $\phi
(h)|_{\rho ^{1}(h)}$, $\phi (j)|_{\rho ^{0}(j)}$, and $\phi (k)|_{\rho
^{2}(k)}$ must be identified on the non-empty pairwise intersections $
U_{\rho ^{0}(j)}\cap U_{\rho ^{1}(h)}$, $U_{\rho ^{1}(h)}\cap U_{\rho
^{2}(k)}$ we can associate to the polytope $|P_{T_{l}}|\rightarrow {M}$ a
complex structure $((M;N_{0}),\mathcal{C})$ by coherently gluing, along the
pattern associated with the ribbon graph $\Gamma $, the local
uniformizations $\{U_{\rho ^{0}(j)}\}_{j=1}^{N_{2}(T)}$, $\{U_{\rho
^{1}(h)}\}_{h=1}^{N_{1}(T)}$, and \ $\{U_{\rho ^{2}(k)}\}_{k=1}^{N_{0}(T)}$.
Explicitly, let $\{U_{\rho ^{1}(j_{\alpha })}\}$, $\alpha =1,2,3$ be the
three generic open strips associated with the three cyclically oriented
edges $\{\rho ^{1}(j_{\alpha })\}$ incident on the generic vertex $\rho
^{0}(j)$. Then the uniformizing coordinates $\{z(j_{\alpha })\}$ are related
to $w(j)$ by the transition functions 
\begin{equation}
w(j)=e^{2\pi i\frac{\alpha -1}{3}}z(j_{\alpha })^{\frac{2}{3}},\hspace{0.2in}\hspace{0.1in}\alpha =1,2,3.  \label{glue1}
\end{equation}
Note that in such uniformization the vertices $\{\rho ^{0}(j)\}$ do not
support conical singularities since each strip $U_{\rho ^{1}(j_{\alpha })}$
is mapped by (\ref{glue1}) into a wedge of angular opening $\frac{2\pi }{3}$. 
This is consistent with the definition of $|P_{T_{l}}|\rightarrow {M}$
according to which the vertices $\{\rho ^{0}(j)\}\in |P_{T_{l}}|\rightarrow {M}$ are the barycenters of the flat $\{\sigma ^{2}(j)\}\in
|T_{l}|\rightarrow M$. Similarly, if $\{U_{\rho ^{1}(k_{\beta })}\}$, $\beta =1,2,...,q(k)$ 
are the open strips associated with the $q(k)$
(oriented) edges $\{\rho ^{1}(k_{\beta })\}$ boundary of the generic
polygonal cell $\rho ^{2}(k)$, then the transition functions between the
corresponding uniformizing coordinate $\zeta (k)$ and the $\{z(k_{\beta })\}$
are given by \cite{mulase}
\begin{equation}
\zeta (k)=\exp \left( \frac{2\pi i}{l(\partial (\rho ^{2}(k)))}\left(
\sum_{\beta =1}^{\nu -1}l(\rho ^{1}(k_{\beta }))+z(k_{\nu })\right) \right) ,\hspace{0.2in}\nu =1,...,q(k),  \label{glue2}
\end{equation}
with $\sum_{\beta =1}^{\nu -1}\cdot \doteq 0$, for $\nu =1$.

\subsection{A Parametrization of the conical geometry.} 
Note that for any closed curve $c:\mathbb{S}^{1}\rightarrow U_{\rho ^{2}(k)}$,
homotopic to the boundary of $\overline{U}_{\rho ^{2}(k)}$, we get 
\begin{equation}
\oint_{c}\sqrt{\phi (k)_{\rho ^{2}(k)}}=l(\partial (\rho ^{2}(k))).
\label{length}
\end{equation}
which shows that the geometry associated with $\phi (k)_{\rho ^{2}(k)}$ is
described by the cylindrical metric canonically associated with a quadratic
differential with a second order pole,\emph{i.e.} 
\begin{equation}
|\phi (k)_{\rho ^{2}(k)}|=\frac{\left[ l(\partial (\rho ^{2}(k)))\right] ^{2}
}{4\pi ^{2}|\zeta (k)|^{2}}|d\zeta (k)|^{2}.  \label{flmetr}
\end{equation}
Thus, as already recalled (section \ref{degenerations}) the punctured disk $\Delta
_{k}^{\ast }\subset U_{\rho ^{2}(k)}$ 
\begin{equation}\label{puncdisk}
\Delta _{k}^{\ast }\doteq \{\zeta (k)\in \mathbb{C}\;|\;0<|\zeta (k)|<1\},
\end{equation}
endowed with the flat metric $|\phi (k)_{\rho ^{2}(k)}|$, is isometric to a
flat semi-infinite cylinder. It is perhaps worthwhile stressing that even
if this latter geometry is perfectly consistent with the metric ribbon graph
structure, it is not really the natural metric to use if we wish to
explicitly keep track of the conical Regge polytope (and the associated
triangulation) which generate the given ribbon graph. For instance, in the
Kontsevich-Witten model, the ribbon graph structure and the associated
geometry $(\Delta _{k}^{\ast },|\phi (k)_{\rho ^{2}(k)}|)$ of the cells 
$U_{\rho ^{2}(k)}$ is sufficient to combinatorially describe intersection
theory on moduli space. However, a full-fledged study of 2D simplicial
quantum gravity (\emph{e.g.}, of Liouville theory) requires a study of the
full Regge geometry $(\Delta _{k}^{\ast },ds_{(k)}^{2})$.

\noindent We can keep track of the conical geometry of the polygonal cell $\rho
^{2}(k)\in $ $|P_{T_{l}}|\rightarrow {M}$ by noticing that for a given
deficit angle $\varepsilon (k)=2\pi -\theta (k)$, the conical geometry 
(\ref{cmetr}) and the cylindrical geometry (\ref{flmetr}) can be
conformally related on each punctured disk $\Delta _{k}^{\ast }\subset
U_{\rho ^{2}(k)}$ according to 
$$ds_{(k)}^{2}\doteq \frac{\left[ l(\partial (\rho ^{2}(k)))\right] ^{2}}{4\pi ^{2}}\left| \zeta (k)\right| ^{-2\left( \frac{\varepsilon (k)}{2\pi }
\right) }\left| d\zeta (k)\right| ^{2}=$$
\begin{equation}\label{metrica} 
=\frac{\left[ l(\partial (\rho ^{2}(k)))\right] ^{2}}{4\pi ^{2}}\left|
\zeta (k)\right| ^{2\left( \frac{\theta (k)}{2\pi }\right) }\frac{\left|
d\zeta (k)\right| ^{2}}{\left| \zeta (k)\right| ^{2}}.  
\end{equation}
It follows that we can apply the explicit construction \cite{mulase} of the mapping
(\ref{bijec}) for defining the decorated Riemann surface corresponding
to a conical Regge polytope (figure 10). Then, an obvious adaptation of theorem 4.2 of
\cite{mulase} provides

\begin{proposition}\label{gluing}
If $\{p_{k}\}_{k=1}^{N_{0}}\in M$ denotes the set of punctures corresponding
to the decorated vertices $\{\sigma ^{0}(k),\frac{\varepsilon (k)}{2\pi }
\}_{k=1}^{N_{0}}$ of the triangulation $|T_{l}|\rightarrow M$, let $RP_{g,N_{0}}^{met}$ be the space of conical Regge polytopes $|P_{T_{l}}|\rightarrow {M}$, with $
N_{0}(T)$ labelled vertices, and let ${\mathfrak{P}}_{g},_{N_{0}}$ be the moduli space ${\mathfrak{M}}_{g},_{N_{0}}$ decorated with the local metric uniformizations 
$(\zeta(k),ds_{(k)}^{2})$  around each puncture $p_k$. Then the map
 $$\Upsilon :RP_{g,N_{0}}^{met}\longrightarrow{\mathfrak{P}}_{g},_{N_{0}}$$ provided by
\begin{gather}
\Upsilon :(|P_{T_{l}}|\rightarrow {M)\longrightarrow }((M;N_{0}),\mathcal{C}
);\{ds_{(k)}^{2}\}) = \label{riemsurf} \\
=\bigcup_{\{\rho ^{0}(j)\}}^{N_{2}(T)}U_{\rho
^{0}(j)}\bigcup_{\{\rho ^{1}(h)\}}^{N_{1}(T)}U_{\rho ^{1}(h)}\bigcup_{\{\rho
^{2}(k)\}}^{N_{0}(T)}(U_{\rho ^{2}(k)},ds_{(k)}^{2}),  \notag
\end{gather}
defines the decorated, $N_{0}$-pointed, Riemann surface $((M;N_{0}),
\mathcal{C})$ canonically associated with the conical Regge polytope $
|P_{T_{l}}|\rightarrow {M}$.
\end{proposition}

\noindent

\noindent In other words, through the morphism $\Upsilon $ defined by the
gluing maps (\ref{glue1}) and (\ref{glue2}) one can generate, from a
polytope barycentrically dual to a (generalized) Regge triangulation, a
Riemann surface $((M;N_{0}),\mathcal{C})\in \overline{\mathfrak{M}}_{g},_{N_{0}}$.

\noindent Such a surface naturally carries the decoration provided by a choice of
local coordinates $\zeta (k)$ around each puncture, of the corresponding
meromorphic quadratic differential $\phi (k)|_{\rho ^{2}(k)}$ (figure 11) and of the
associated conical metric $ds_{(k)}^{2}$. It is through such a decoration
that the punctured Riemann surface $((M;N_{0}),\mathcal{C})$ keeps track of
the metric geometry of the conical Regge polytope $|P_{T_{l}}|\rightarrow {M}
$ out of which $((M;N_{0}),\mathcal{C})$\ has been generated.  

\subsection{Opening of the cones and deformations of the Regge
geometry.} 
According to the procedure described in section \ref{conical sectors}, we can
open the cones $(U_{\rho ^{2}(k)},ds_{(k)}^{2})=\cup _{\alpha
=1}^{q(k)}S_{\alpha }(k)$ and map each conical sector $S_{\alpha }(k)$ onto
a corresponding annulus 
\begin{equation}\label{ann2}
\Delta _{\vartheta _{\alpha }(k)}(k)\doteq \{W(k)\in \mathbb{C}\; |\;e^{-2\pi
\vartheta _{\alpha }(k)}<\left| W(k)\right| <1\},  
\end{equation}
(see(\ref{annulus}),(\ref{modulann})). In such a setting, a natural question
to discuss concerns how a deformation of the conical sectors $S_{\alpha }(k)$
, at fixed deficit angle $\varepsilon (k)$ (and perimeter $l(\partial (\rho
^{2}(k)))$), affects the conical geometry of $(U_{\rho ^{2}(k)},ds_{(k)}^{2})
$, and how this deformation propagates to the underlying Regge polytope $
|P_{T_{l}}|\rightarrow {M}$. The question is analogous to the study of the
non-trivial deformations of constant curvature metrics on a surface. To
this end, we adapt to our purposes a standard procedure by considering the
following deformation of $S_{\alpha }(k)$ 
\begin{equation}
\vartheta _{\alpha }(k)\longmapsto \vartheta _{\alpha }(k)^{\prime }\doteq
(s+1)\vartheta _{\alpha }(k),\;s\in \mathbb{R},
\end{equation}
and discuss its effect on the map $\Upsilon $ near the identity $s=0$. In
the annulus description of the geometry of the corresponding sector $
S_{\alpha }(k)$, such a deformation is realized by the quasi-conformal map 
\begin{equation}
W(k)\longmapsto f(W(k))=W(k)\left| W(k)\right| ^{s},
\end{equation}
which indeed maps the annulus (\ref{ann2}) into the annulus 
\begin{equation}
\{W(k)\in \mathbb{C}\; |\;e^{-2\pi (s+1)\vartheta _{\alpha }(k)}<\left|
W(k)\right| <1\},
\end{equation}
corresponding to a conical sector of angle $\vartheta _{\alpha }(k)^{\prime
}\doteq (s+1)\vartheta _{\alpha }(k)$. The study of the transformation $
f(W(k))$ is well known in the modular theory of the annulus (see \cite{wolpert}), and goes as follows. The Beltrami differential associated
with $f(\zeta (k))$ at $s=0$, is given by 
\begin{equation}
\frac{\partial }{\partial \overline{W(k)}}\left[ \left. \frac{\partial }{
\partial s}f(W(k))\right| _{s=0}\right] =\frac{W(k)}{2\overline{W(k)}}.
\end{equation}
The corresponding Beltrami differential on (\ref{ann2}) representing the
infinitesimal deformation of $S_{\alpha }(k)$ in the direction $\partial
/\partial (e^{-2\pi \vartheta _{\alpha }(k)})$ is provided by 
\begin{eqnarray}
\mu _{\alpha }(k) &\doteq &\left( \frac{\partial e^{-2\pi (s+1)\vartheta
_{\alpha }(k)}}{\partial s}\right) _{s=0}^{-1}\frac{\partial }{\partial 
\overline{W(k)}}\left[ \left. \frac{\partial }{\partial s}f(W(k))\right|
_{s=0}\right] = \\
&=&-\left( \frac{1}{2\pi \vartheta _{\alpha }(k)}\right) e^{2\pi \vartheta
_{\alpha }(k)}\frac{W(k)}{2\overline{W(k)}}.  \notag
\end{eqnarray}

\noindent Recall that a Beltrami differential on the annulus is naturally paired with
a quadratic differential $\widehat{\phi _{\alpha }(k)}\doteq C_{k}\frac{
dW(k)\otimes dW(k)}{W^{2}(k)}$, $C_{k}$ being a real constant, via the $L^{2}
$ inner product 
\begin{equation}
\left\langle \mu _{\alpha }(k),\widehat{\phi _{\alpha }(k)}\right\rangle
_{\Delta _{\vartheta _{\alpha }(k)}(k)}\doteq \frac{\sqrt{-1}}{2}
\int_{\Delta _{\vartheta _{\alpha }(k)}(k)}\mu (k)\widehat{\phi _{\alpha }(k)
}dW(k)\wedge d\overline{W(k)}.
\end{equation}
By requiring that 
\begin{equation}
\left\langle \mu _{\alpha }(k),\widehat{\phi _{\alpha }(k)}\right\rangle
_{\Delta _{\vartheta _{\alpha }(k)}(k)}=1
\end{equation}
one finds the constant $C_{k}$ characterizing the quadratic differential $
\widehat{\phi _{\alpha }(k)}$, dual to $\mu _{\alpha }(k)$, and describing
the infinitesimal deformation of $S_{\alpha }(k)$ in the direction $
d(e^{-2\pi \vartheta _{\alpha }(k)})$, (\emph{i.e.}, a cotangent vector to $
\mathfrak{M}_{g},_{N_{0}}$ at $((M;N_{0}),\mathcal{C});\{ds_{(k)}^{2}\})$). 
A direct computation provides 
\begin{equation}
\widehat{\phi _{\alpha }(k)}=-\frac{e^{-2\pi \vartheta _{\alpha }(k)}}{\pi }
\frac{dW(k)\otimes dW(k)}{W^{2}(k)}.
\end{equation}
Note that whereas $\phi (k)_{\rho ^{2}(k)}$ may be thought of as a cotangent
vector to one of the $\mathbb{R}^{N_{0}}$ fiber of $\mathfrak{M}
_{g},_{N_{0}}\times \mathbb{R}_{+}^{N}$, the quadratic differential $
\widehat{\phi _{\alpha }(k)}$ projects down to a deformation in the
base $\mathfrak{M}_{g},_{N_{0}}$, and as such is a much more interesting object
than $\phi (k)_{\rho ^{2}(k)}$. The modular nature of $\widehat{\phi
_{\alpha }(k)}$ comes clearly to the fore if we compute its associated
Weil-Petersson norm according to 
\begin{equation}
\left\| \widehat{\phi _{\alpha }(k)}\right\| _{W-P}=\int_{\Delta _{\vartheta
_{\alpha }(k)}(k)}\frac{|\widehat{\phi _{\alpha }(k)}|^{2}}{g_{hyp}},
\end{equation}
where $g_{hyp}$ denotes the hyperbolic metric 
\begin{equation}
g_{hyp}\doteq \left( \frac{\pi ^{2}}{\ln ^{2}|t_{\alpha }(k)|}\right) \frac{
dW(k)d\overline{W(k)}}{|W(k)|^{2}\sin ^{2}\left( \pi \frac{\ln |W(k)|}{\ln
|t_{\alpha }(k)|}\right) }
\end{equation}
on the annulus $\Delta _{\vartheta _{\alpha }(k)}(k)$, and where $|t_{\alpha
}(k)|\doteq e^{-2\pi \vartheta _{\alpha }(k)}$. A direct computation
provides \cite{wolpert}
\begin{equation}
\left\| \widehat{\phi _{\alpha }(k)}\right\| _{W-P}=\left| t_{\alpha
}(k)\right| ^{2}\left( \frac{1}{\pi }\ln \frac{1}{|t_{\alpha }(k)|}\right)
^{3}=8\vartheta _{\alpha }^{3}(k)e^{-4\pi \vartheta _{\alpha }(k)}.
\label{WPnomr}
\end{equation}
According to proposition \ref{gluing}, the decorated Riemann surface 
$$(((M;N_{0}),\mathcal{C});\{ds_{(k)}^{2}\})$$
is generated by glueing the oriented domains $(U_{\rho ^{2}(k)},ds_{(k)}^{2})$ along the ribbon graph $\Gamma $
associated with the Regge polytope $|P_{T_{l}}|\rightarrow {M}$. We can indipendently and holomorphically open
the distinct cones $\{(U_{\rho ^{2}(k)},ds_{(k)}^{2})\}$ on corresponding
sectors $\{S_{\alpha }(k)\}$, then it follows that to any such $(((M;N_{0}),
\mathcal{C});\{ds_{(k)}^{2}\})$ we can associate a well defined
Weil-Petersson metric which can be immediately read off from (\ref{WPnomr}).
Since there are $6g-6+2N_{0}$ independent conical sectors $S_{\alpha }(k)$
on the Riemann surface $(((M;N_{0}),\mathcal{C});\{ds_{(k)}^{2}\})$
corresponding to the Regge polytope $|P_{T_{l}}|\rightarrow {M}$, (at fixed
deficit angles $\{\varepsilon (k)\}$ and perimeters $\{l(\partial (\rho
^{2}(k)))\}$), we can find a corresponding sequence of \ angles $\{\vartheta
_{\alpha }(k)\}$ and relabel them as $\{\vartheta _{H}\}_{H=1}^{6g-6+2N_{0}}$. 
If we define 
\begin{equation}
\tau _{H}\doteq e^{-2\pi \vartheta _{H}}
\end{equation}
then we can immediately write the Weil-Petersson metric $ds_{W-P}^{2}$
associated with the Regge polytope $|P_{T_{l}}|\rightarrow {M}$ 
\begin{eqnarray}
ds_{W-P}^{2}(|P_{T_{l}}|) &=&\sum_{H=1}^{6g-6+2N_{0}}\frac{2\pi ^{3}d\tau
_{H}^{2}}{\tau _{H}^{2}\left( \ln \frac{1}{\tau _{H}}\right) ^{3}}=
 \notag \\
&=&\pi ^{2}\sum_{H=1}^{6g-6+2N_{0}}\frac{d\vartheta _{H}^{2}}{\vartheta
_{H}^{3}}.  \label{WPmetr1} 
\end{eqnarray}
If we respectively denote by $\theta (H)$ and $l(H)$ the (fixed) conical
angle $\theta (k)$ and the perimeter $l(\partial (\rho ^{2}(k)))$
corresponding to the modular variable $\vartheta _{H}$, (note that the same $
\theta (H)$ and $l(H)$ correspond to the distinct $\{\vartheta _{H}\}$ which
are incident on the same vertex), then according to (\ref{angles}) we can
rewrite (\ref{WPmetr1}) as 
\begin{equation}
ds_{W-P}^{2}(|P_{T_{l}}|)=\sum_{H=1}^{6g-6+2N_{0}}\frac{\pi ^{2}}{\theta (H)}
\left( \frac{\widehat{L}_{H}}{l(H)}\right) ^{-3}d\left( \frac{\widehat{L}_{H}
}{l(H)}\right) \otimes d\left( \frac{\widehat{L}_{H}}{l(H)}\right) ,
\label{WPmetr2}
\end{equation}
where $\widehat{L}_{H}$ is the length variable $\widehat{L}_{\alpha }(k)$
associated with $\vartheta _{H}$.

Note that (\ref{WPmetr1}) and (\ref{WPmetr2})
show a singular behavior when $\vartheta _{H}\rightarrow 0$ (or equivalently
when $\widehat{L}_{H}\rightarrow 0$), such a behavior occurs for instance
when the Regge triangulation (and the corresponding polytope) degenerates
and exhibits vertices where thinner and thinner triangles are incident. As
recalled in paragraph \ref{degenerations}, this is usually considered a pathology of Regge
calculus. However, in view of the modular correspondence we have  described
in this paper, it simply corresponds to the well-known incompleteness (with
respect of the complex structure of the moduli space) of the Weil-Petersson
metric on $\mathfrak{M}_{g},_{N_{0}}$ as we approach the boundary (the
compactifying divisor) of $\mathfrak{M}_{g},_{N_{0}}$ in $\overline{\mathfrak{M}}
_{g},_{N_{0}}$ . In such a sense, as already remarked in the introductory
remarks, such a pathology of Regge triangulations has a modular meaning and
is not accidental.

\subsection{The Weil-Petersson measure on the space of
random Regge polytopes.}
With the metric $ds_{W-P}^{2}(|P_{T_{l}}|)$ we can associate a well defined volume form 
\begin{gather}
\Omega _{W-P}(|P_{T_{l}}|)= \\
=\left[ \det \left( \frac{\pi ^{2}}{\theta (H)}\left( \frac{\widehat{L}_{H}}{
l(H)}\right) ^{-3}\right) \right] ^{\frac{1}{2}}d\left( \frac{\widehat{L}_{1}
}{l(1)}\right) \bigwedge ...\bigwedge d\left( \frac{\widehat{L}_{6g-6+2N_{0}}
}{l(6g-6+2N_{0})}\right) ,  \notag
\end{gather}
which can be rewritten as a power of product of
$2$-forms of the type $d\left( \frac{\widehat{L}_{H}}{l(H)}\right) \wedge
d\left( \frac{\widehat{L}_{H+1}}{l(H+1)}\right) $, (this being connected
with the K\"{a}hler form associated with (\ref{WPmetr2})). Such forms are
directly related with the Chern classes (\ref{chern}) of the line bundles $
\mathcal{CL}_{k}$ and play a distinguished role in the Kontsevich-Witten
model. The possibility of expressing the Weil-Petersson K\"{a}hler form in
terms of the Chern classes (\ref{chern}) is a deep and basic fact of the
geometry of $\overline{\mathfrak{M}}_{g},_{N_{0}}$, and it is a pleasant feature
of the model discussed here that such a connection can be motivated by
rather elementary considerations.

The Weil-Petersson volume form $\Omega
_{W-P}(|P_{T_{l}}|)$ allows us to integrate over the space 
\begin{equation}
RP_{g,N_{0}}^{met}(\{\varepsilon (H)\},\{l(H)\})
\end{equation}
of distinct Regge polytopes $|P_{T_{l}}|\rightarrow {M}$ with given deficit
angles $\{\varepsilon (k)\}$ and perimeters $\{l(\partial (\rho ^{2}(k)))\}$.  
To put such an integration in a proper perspective and give it a
suggestive physical meaning we shall explicitly consider the set of Regge
polytopes 
\begin{equation}\label{RPg}
RP_{g,N_{0}}^{met}(\{q(H)\})\doteq \left\{ |P_{T_{l}}|\rightarrow {M|\;}
\varepsilon (H)=2\pi -\frac{\pi }{3}q(H);\;l(H)=\frac{\sqrt{3}}{3}
aq(H)\right\} ,
\end{equation}
which contain the equilateral Regge polytopes dual to dynamical
triangulations. According to proposition \ref{gluing}, $RP_{g,N_{0}}^{met}(\{q(H)\})$
is a combinatorial description of $\mathfrak{M}_{g},_{N_{0}}$, and the volume
form $\Omega _{W-P}(|P_{T_{l}}|)$ can be considered as the pull-back under
the morphism $\Upsilon $, (see (\ref{riemsurf})) of the Weil-Petersson
volume form $\omega _{WP}^{3g-3+N_{0}}/(3g-3+N_{0})!$ on $\mathfrak{M}
_{g},_{N_{0}}$. 

Stated differently, the integration of $\Omega
_{W-P}(|P_{T_{l}}|)$ over the space (\ref{RPg}) is
equivalent to the Weil-Petersson volume of $\mathfrak{M}_{g},_{N_{0}}$. It is
well-known that the Weil-Petersson form $\omega _{WP}$ extends (as a (1,1)
current) to the compactification $\overline{\mathfrak{M}}_{g},_{N_{0}}$, and, if
we denote by $\overline{RP}_{g,N_{0}}^{met}(\{q(H)\})$ the compactified
orbifold associated with $RP_{g,N_{0}}^{met}(\{q(H)\})$, then we can write 
$$\frac{1}{N_{0}!}\int_{\overline{RP}_{g,N_{0}}^{met}(\{q(H)\})}\Omega
_{W-P}(|P_{T_{l}}|)=$$
\begin{equation}\label{integ} 
=\frac{1}{N_{0}!}\int_{\overline{\mathfrak{M}}_{g},_{N_{0}}}\frac{\omega
_{WP}^{3g-3+N_{0}}}{(3g-3+N_{0})!}=Vol\left( \overline{\mathfrak{M}}
_{g},_{N_{0}}\right)  
\end{equation}
where we have divided by $N_{0}(T)!$ in order to factor out the labelling of
the $N_{0}(T)$ punctures. Since $\overline{RP}_{g,N_{0}}^{met}(\{q(H)\})$
is a (smooth) stratified orbifold acted upon by the automorphism group $
Aut_{\partial }(P_{T_{a}})$, we can explicitly write the left side of (\ref
{integ}) as an orbifold integration over the distinct orbicells $\Omega
_{T_{a}}$ (\ref{strata}) in which $\overline{RP}_{g,N_{0}}^{met}(\{q(H)\})$
is stratified

\begin{gather}
\frac{1}{N_{0}!}\sum_{T\in \mathcal{DT}[\{q(i)\}_{i=1}^{N_{0}}]}\frac{1}{
|Aut_{\partial }(P_{T_{a}})|}\int_{_{\Omega
_{T_{a}}(\{q(k)\}_{k=1}^{N_{0}})}}\Omega _{W-P}(|P_{T_{l}}|)= \notag 
\\
\notag \\
=VOL\left( \overline{\mathfrak{M}}_{g},_{N_{0}}\right),  \label{orbin}
\end{gather}
(the orbifold integration is defined in \cite{penner}, Th. 3.2.1), where the
summation is over all distinct dynamical triangulations $\mathcal{DT}[\left\{q(i)\right\}_{i=1}^{N_0}]$ with given unlabeled
curvature assignments weighted by the order $|Aut_{\partial }(P_{T})|$ of
the automorphisms group of the corresponding dual polytope. The relation (
\ref{orbin}) provides a non-trivial connection between dynamical
triangulations (labelling the strata of $\overline{RP}_{g,N_{0}}^{met}(
\{q(H)\})$ (or, equivalently, of $\overline{\mathfrak{M}}_{g},_{N_{0}}$), and
the fixed connectivity Regge triangulations in each strata $\Omega _{T_{a}}$. 
Some aspects of this relation have already been discussed by us elsewhere,
\cite{carfora}. Here we will exploit (\ref{orbin}
) in order to directly relate $VOL\left( \overline{\mathfrak{M}}
_{g},_{N_{0}}\right) $ to the partition function of 2D simplicial quantum
gravity. To this end let us sum both members of (\ref{orbin}) over the set
of all possible curvature assignments $\{q(H)\}$ on the $N_{0}$ unlabelled
vertices of the triangulations, and note that 
\begin{multline}
Card\left[ \mathcal{DT}(N_{0})\right] \doteq \sum_{\mathcal{DT}(N_{0})}\frac{
1}{|Aut_{\partial }(P_{T_{a}})|}= \notag \\
=\frac{1}{N_{0}!}\sum_{\{q(H)\}_{H=1}^{N_{0}}}\sum_{T\in \mathcal{DT}
[\{q(H)\}_{H=1}^{N_{0}}]}\frac{1}{|Aut_{\partial }(P_{T_{a}})|}  
\end{multline}
provides the number of distinct (generalized) dynamical triangulations with $
N_{0}$ unlabelled vertices. Since $VOL\left( \overline{\mathfrak{M}}
_{g},_{N_{0}}\right) $ does not depend of the curvature assignments $\{q(H)\}
$, from (\ref{orbin}) we get 
\begin{gather}
\frac{1}{N_{0}!}\sum_{\{q(H)\}_{H=1}^{N_{0}}}\sum_{T\in \mathcal{DT}
[\{q(i)\}_{i=1}^{N_{0}}]}\frac{1}{|Aut_{\partial }(P_{T_{a}})|}
\int_{_{\Omega _{T_{a}}(\{q(k)\}_{k=1}^{N_{0}})}}\Omega _{W-P}(|P_{T_{l}}|)=
 \notag \\
=\sum_{\mathcal{DT}(N_{0})}\frac{1}{|Aut_{\partial }(P_{T_{a}})|}
\int_{_{\Omega _{T_{a}}(\{q(k)\}_{k=1}^{N_{0}})}}\Omega _{W-P}(|P_{T_{l}}|)=
\notag \\
=\left( Card\{q(H)\}\right) VOL\left( \overline{\mathfrak{M}}_{g},_{N_{0}}
\right), 
\end{gather}
where $Card\{q(H)\}$ denotes the number of possible curvature assignments 
on the $N_{0}$ unlabelled vertices of the triangulations. Dividing
both members by $\left(Card\{q(H)\}\right) $, we eventually get the relation
\begin{equation}
\sum_{\mathcal{DT}(N_{0})}\frac{1}{|Aut_{\partial }(P_{T_{a}})|}
\int_{_{\Omega _{T_{a}}(\{q(k)\}_{k=1}^{N_{0}})}}\frac{\Omega
_{W-P}(|P_{T_{l}}|)}{Card\{q(H)\}}=VOL\left( \overline{\mathfrak{M}}
_{g},_{N_{0}}\right) ,  \label{partfunc}
\end{equation}
(the number $Card\{q(H)\}$ has been shifted under the integral sign for
typographical convenience). We have the following 

\begin{lemma}
For $N_{0}$ sufficiently large there exist a positive constant $K(g)$
independent from $N_{0}$ but possibly dependent on the genus $g$, and a
positive constant $\kappa $ independent both from $N_{0}$ and $g$ such that
\begin{equation}
\int_{_{\Omega _{T_{a}}(\{q(k)\}_{k=1}^{N_{0}})}}\frac{\Omega
_{W-P}(|P_{T_{l}}|)}{Card\{q(H)\}}\simeq K(g)e^{-\kappa N_{0}}.
\end{equation}
\end{lemma}

\bigskip 

In order to prove this result let us start by recalling that the large $
N_{0}(T)$ asymptotics of the triangulation counting $Card\left[ \mathcal{DT}
[N_{0}]\right] $ can be obtained from purely combinatorial (and matrix
theory) arguments \cite{mulase}, \cite{brezin} to the effect that 
\begin{equation}
Card\left[ \mathcal{DT}[N_{0}]\right] \sim \frac{16c_{g}}{3\sqrt{2\pi }}
\cdot e^{\mu _{0}N_{0}(T)}N_{0}(T)^{\frac{5g-7}{2}}\left( 1+O(\frac{1}{N_{0}}
)\right) ,  \label{oldentr}
\end{equation}
where $c_{g}$ is a numerical constant depending only on the genus $g$, and $
e^{\mu _{0}}=(108\sqrt{3})$ is a (non-universal) parameter depending on the
set of triangulations considered (here the generalized triangulations,
barycentrically dual to trivalent graphs; in the case of regular
triangulations in place of $108\sqrt{3}$ we would get $e^{\mu _{0}}=(\frac{
4^{4}}{3^{3}})$). Thus, if we denote by 
\begin{gather}
\left\langle \int \frac{\Omega _{W-P}(|P_{T_{l}}|)}{Card\{q(H)\}}
\right\rangle _{\mathcal{DT}[N_{0}]}\doteq \notag \\
=\frac{1}{Card\left[ \mathcal{DT}[N_{0}]\right] }\sum_{\mathcal{DT}(N_{0})}
\frac{1}{|Aut_{\partial }(P_{T_{a}})|}\int_{_{\Omega
_{T_{a}}(\{q(k)\}_{k=1}^{N_{0}})}}\frac{\Omega _{W-P}(|P_{T_{l}}|)}{
Card\{q(H)\}}  
\end{gather}
the average value of $\int $\ $\Omega _{W-P}(|P_{T_{l}}|)/Card\{q(H)\}$
over the set $\mathcal{DT}[N_{0}]$, (dropping the integration range $\Omega
_{T_{a}}(\{q(k)\}_{k=1}^{N_{0}})$ for notational ease),  then we can write
the large $N_{0}$ asymptotics of the left side member of (\ref{orbin}) as
\begin{gather}
\sum_{\mathcal{DT}(N_{0})}\frac{1}{|Aut_{\partial }(P_{T_{a}})|}
\int_{_{\Omega _{T_{a}}(\{q(k)\}_{k=1}^{N_{0}})}}\frac{\Omega
_{W-P}(|P_{T_{l}}|)}{Card\{q(H)\}}\simeq   \label{aver}  \notag \\
\simeq \frac{16c_{g}}{3\sqrt{2\pi }}\left\langle \int \frac{\Omega
_{W-P}(|P_{T_{l}}|)}{Card\{q(H)\}}\right\rangle _{\mathcal{DT}[N_{0}]}e^{\mu
_{0}N_{0}(T)}N_{0}(T)^{\frac{5g-7}{2}}\left( 1+O(\frac{1}{N_{0}})\right). 
\end{gather}
On the other hand, from the Manin-Zograf asymptotic analysis of 
$VOL\left( \overline{\mathfrak{M}}_{g},_{N_{0}}\right) $ for fixed genus $g$ and large 
$N_{0}$, we have\cite{manin}\cite{zograf} 
\begin{gather}
VOL\left( \overline{\mathfrak{M}}_{g},_{N_{0}}\right) =   \notag \\
=\pi ^{2(3g-3+N_{0})}(N_{0}+1)^{\frac{5g-7}{2}}C^{-N_{0}}\left(
B_{g}+\sum_{k=1}^{\infty }\frac{B_{g,k}}{(N_{0}+1)^{k}}\right),  \label{manzog}
\end{gather}
where $C=-\frac{1}{2}j_{0}\frac{d}{dz}J_{0}(z)|_{z=j_{0}}$, ($J_{0}(z)$ the

Bessel function, $j_{0}$ its first positive zero); (note that $C\simeq
0.625....$). The genus dependent parameters $B_{g}$ are explicitly given
\cite{zograf} by 
\begin{equation}
\left\{ 
\begin{tabular}{ll}
$B_{0}=\frac{1}{A^{1/2}\Gamma (-\frac{1}{2})C^{1/2}},$ & $B_{1}=\frac{1}{48},
$ \\ 
$B_{g}=\frac{A^{\frac{g-1}{2}}}{2^{2g-2}(3g-3)!\Gamma (\frac{5g-5}{2})C^{
\frac{5g-5}{2}}}\left\langle \tau _{2}^{3g-3}\right\rangle ,$ & $g\geq 2$
\end{tabular}
\right. 
\end{equation}
where $A\doteq -j_{0}^{-1}J_{0}^{\prime }(j_{0})$, and $\left\langle \tau
_{2}^{3g-3}\right\rangle $ is a Kontsevich-Witten intersection number \cite{kontsevich}\cite{witten}, (the
coefficients $B_{g,k}$ can be computed similarly-see \cite{zograf} for details). By
Inserting (\ref{manzog}) in the left hand side of (\ref{orbin}) and by
taking into account (\ref{aver}) we eventually get 
\begin{gather}
\frac{16c_{g}}{3\sqrt{2\pi }}\left\langle \int \frac{\Omega
_{W-P}(|P_{T_{l}}|)}{Card\{q(H)\}}\right\rangle _{\mathcal{DT}[N_{0}]}e^{\mu
_{0}N_{0}(T)}N_{0}(T)^{\frac{5g-7}{2}}\left( 1+O(\frac{1}{N_{0}})\right) = \\
=\pi ^{2(3g-3+N_{0})}(N_{0}+1)^{\frac{5g-7}{2}}C^{-N_{0}}\left(
B_{g}+\sum_{k=1}^{\infty }\frac{B_{g,k}}{(N_{0}+1)^{k}}\right) ,  \notag
\end{gather}
which by direct comparison provides
\begin{equation}
\left\langle \int \frac{\Omega _{W-P}(|P_{T_{l}}|)}{Card\{q(H)\}}
\right\rangle _{\mathcal{DT}[N_{0}]}\simeq \frac{3\sqrt{2\pi }B_{g}\pi
^{2(3g-3+N_{0})}}{16c_{g}}e^{(|\ln C|-\mu _{0})N_{0}}.
\end{equation}
This proves the lemma with 
\begin{equation}
K(g)=\frac{3\sqrt{2\pi }B_{g}\pi ^{2(3g-3)}}{16c_{g}},
\end{equation}
and 
\begin{equation}
\kappa =\mu _{0}-|\ln C|-2\ln\pi\approx 2.472.  \label{kappa}
\end{equation}
Thus, according to the above lemma, we can rewrite the left member of (\ref
{partfunc}) as
\begin{gather}
VOL\left( \overline{\mathfrak{M}}_{g},_{N_{0}}\right) =\sum_{\mathcal{DT}(N_{0})}
\frac{1}{|Aut_{\partial }(P_{T_{a}})|}\int_{_{\Omega
_{T_{a}}(\{q(k)\}_{k=1}^{N_{0}})}}\frac{\Omega _{W-P}(|P_{T_{l}}|)}{
Card\{q(H)\}}=  \label{DTpart} \\
=K(g)\sum_{\mathcal{DT}(N_{0})}\frac{1}{|Aut_{\partial }(P_{T_{a}})|}
e^{-\kappa N_{0}},  \notag
\end{gather}
which has the structure of the canonical partition function for dynamical
triangulation theory \cite{ambjorn}. Roughly speaking, we may interpret (\ref{DTpart})
by saying that dynamical triangulations count the (orbi)cells which
contribute to the volume (each cell containing the Regge triangulations
whose adjacency matrix is that one of the dynamical triangulation which
label the cell), whereas integration over the cells weights the fluctuations
due to all Regge triangulations which, in a sense, represent the
deformational degrees of freedom of the given dynamical triangulation
labelling the cell.

\bigskip 

\section{Concluding remarks}

Note that in the canonical partition function of 2D simplicial quantum
gravity, $-\kappa $ has the role of a running coupling constant (a chemical
potential in a grand-canonical description), whereas in (\ref{DTpart}) it
takes on the value (\ref{kappa}). This remark suggests that $e^{\lambda
N_{0}}VOL\left( \overline{\mathfrak{M}}_{g},_{N_{0}}\right) $ with $\lambda $ a
running coupling constant is the canonical partition function of 2D
simplicial quantum gravity. More generally, if we let also fluctuate the
topology of the triangulated surfaces $|T_{l}|\rightarrow M$ then we have
strong elements suggesting that the canonical partition function of 2D
simplicial quantum gravity over the set of (orientable) surfaces with random
topology  is 
\begin{equation}
e^{\lambda N_{0}+\eta g}VOL\left( \overline{\mathfrak{M}}_{g},_{N_{0}}\right) .
\end{equation}
In this connection it  is also important to stress that (\ref{manzog}) shows
very naturally how the genus-$g$ pure gravity critical exponent

\begin{equation}
\gamma _{g}=\frac{5g-1}{2}.
\end{equation}
originates from the cardinality of the cell decomposition of $\overline{
\mathfrak{M}}_{g},_{N_{0}}$ parametrized by $T\in \mathcal{DT}
[\{q(i)\}_{i=1}^{N_{0}}]$, (this remark also explains geometrically why
standard, fixed connectivity Regge calculus, is not able to recover $\gamma
_{g}$). 

As already argued in \cite{troyanov} on a slightly different basis, we have a natural
candidate which should play in the matter setting the same role of 
$\overline{\mathfrak{M}}_{g},_{N_{0}}$ in the pure gravity case. This is the
space $\overline{\mathfrak{M}}_{g},_{N_{0}}(X,\beta )$ of stable maps from 
$((M;N_{0}),\mathcal{C})$ to the manifold $X$ parametrizing the matter
fields and representing the homology class $\beta \in H_{2}(X,\mathbb{Z})$,
(for instance, in the case of the Polyakov action, the map can be thought of
as providing the embedding of $((M;N_{0}),\mathcal{C})$ in the target
(Euclidean) space). According to the above remarks, it is rather natural to
conjecture that  the canonical partition function describing conformal
matter interacting with 2D simplicial quantum gravity is provided by a
(generalized) Weil-Petersson volume of $\overline{\mathfrak{M}}
_{g},_{N_{0}}(X,\beta )$. Candidates for such volumes are related to the
(descendent) Gromov-Witten invariants of the manifold $X$, and evidence in
such a direction comes from the well-known fact that many relevant
properties of $\overline{\mathfrak{M}}_{g},_{N_{0}}(X,\beta )$ are governed, via
its intersection theory, by matrix models \cite{okounkov}\cite{eguchi}.
Bringing such a program
to completion appear a rather formidable task, however it will certainly
throw new light on the deep nature of simplicial techniques, Regge calculus
and their foundational role in disclosing the physics of quantum gravity.

\section*{Acknowledgements}

This work was supported in part by the Ministero dell'Universita' e della
Ricerca Scientifica under the PRIN project \emph{The geometry of integrable
systems.}

\newpage

\begin{figure}[ht]
\begin{center}
\includegraphics[bb= 0 0 580 620,scale=.5]{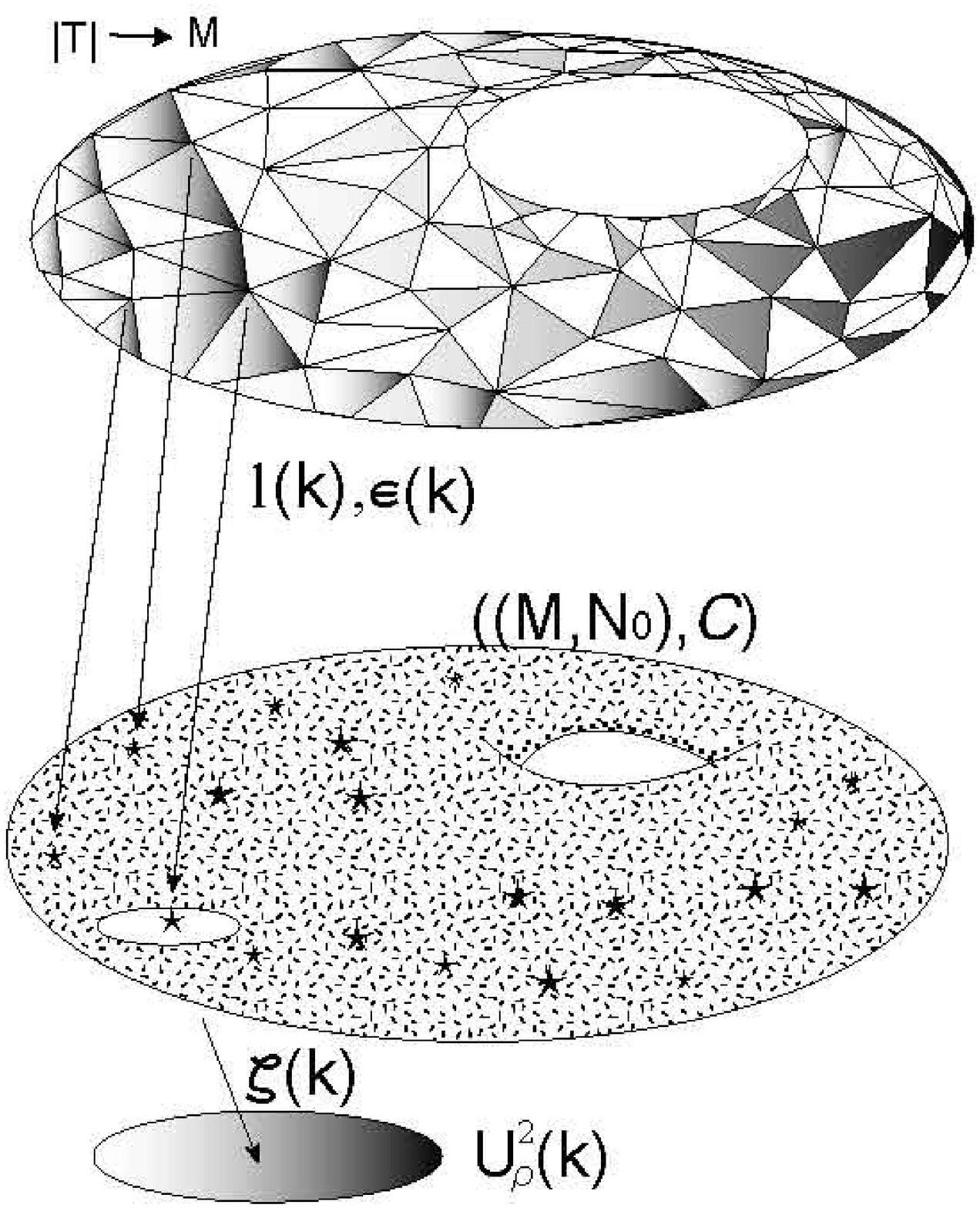}
\caption{A torus triangulated with triangles of variable edge-length.}\label{fig1}
\end{center}
\end{figure}

\newpage

\begin{figure}[ht] 
\begin{center}
\includegraphics[bb= 0 0 650 770,scale=.5]{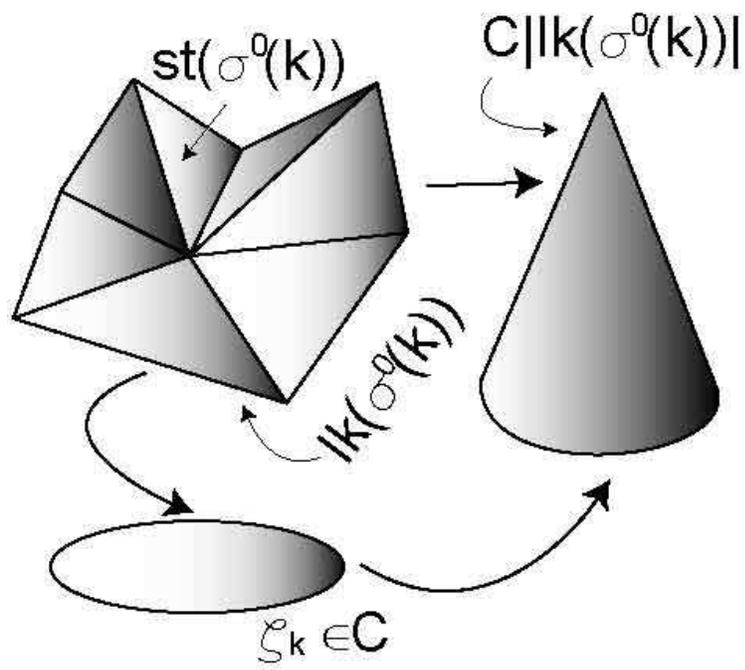}
\caption{The geometric structures around a vertex.}\label{fig2}
\end{center}
\end{figure}

\newpage

\begin{figure}[ht]
\begin{center}
\includegraphics[bb= 0 0 330 830,scale=.5]{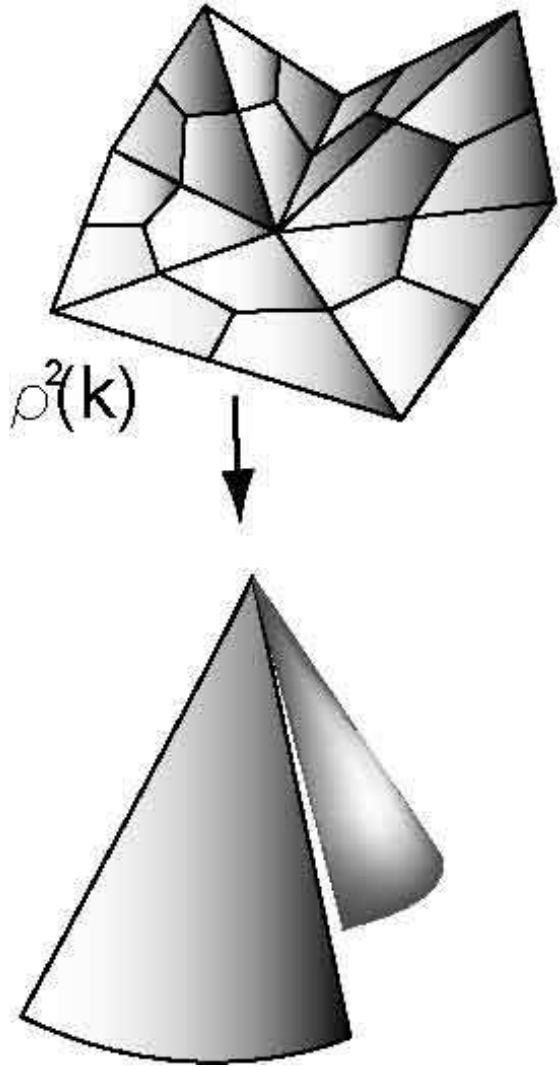}
\caption{The conical geometry of the baricentrically dual polytope.}\label{fig3}
\end{center}
\end{figure}

\newpage

\begin{figure}[ht]
\begin{center}
\includegraphics[bb= 0 0 450 580,scale=.6]{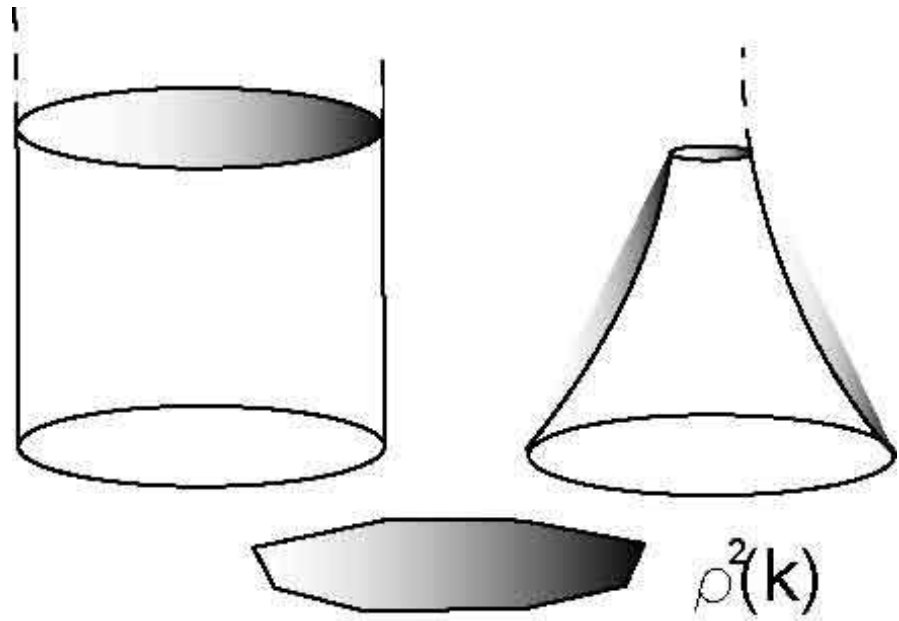}
\caption{The cylindrical and hyperbolic metric over a $\theta\to 0$ degenerating polytopal cell.}\label{fig4}
\end{center}
\end{figure}

\newpage

\begin{figure}[ht]
\begin{center}
\includegraphics[bb= 0 0 550 710,scale=.6]{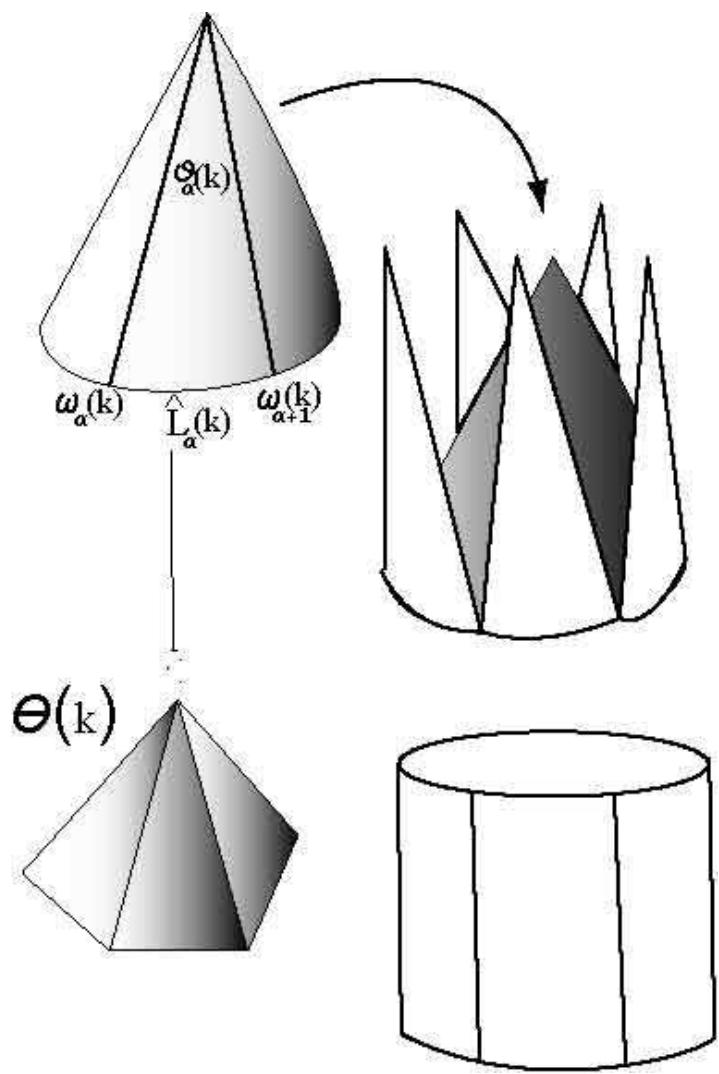}
\caption{The opening of the cone into its constituents conical sectors and the associated cylindrical strip.}\label{fig5}
\end{center}
\end{figure}

\newpage

\begin{figure}[ht]
\begin{center}
\includegraphics[bb= 70 0 470 720,scale=.6]{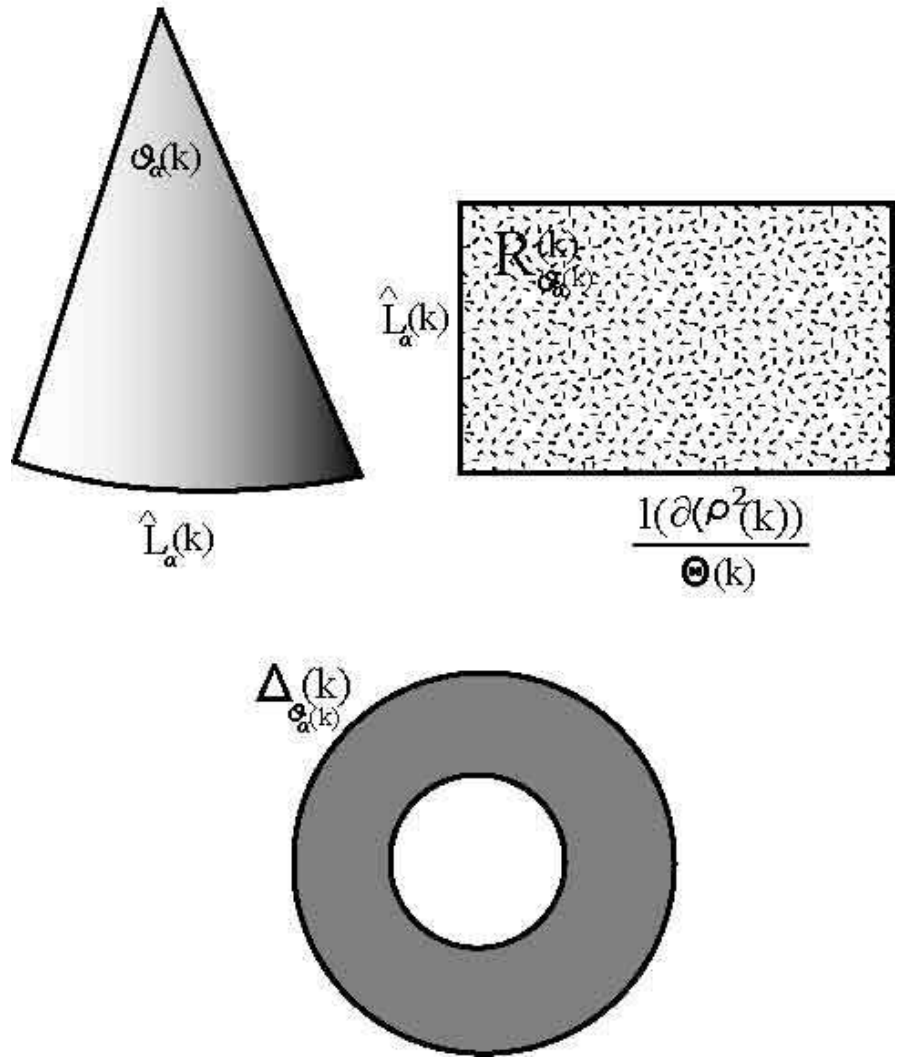}
\caption{The mapping of a conical sector in a strip and into an anular
region.}\label{fig6}
\end{center}
\end{figure}

\newpage

\begin{figure}[ht]
\begin{center}
\includegraphics[bb= 0 0 450 840,scale=.6]{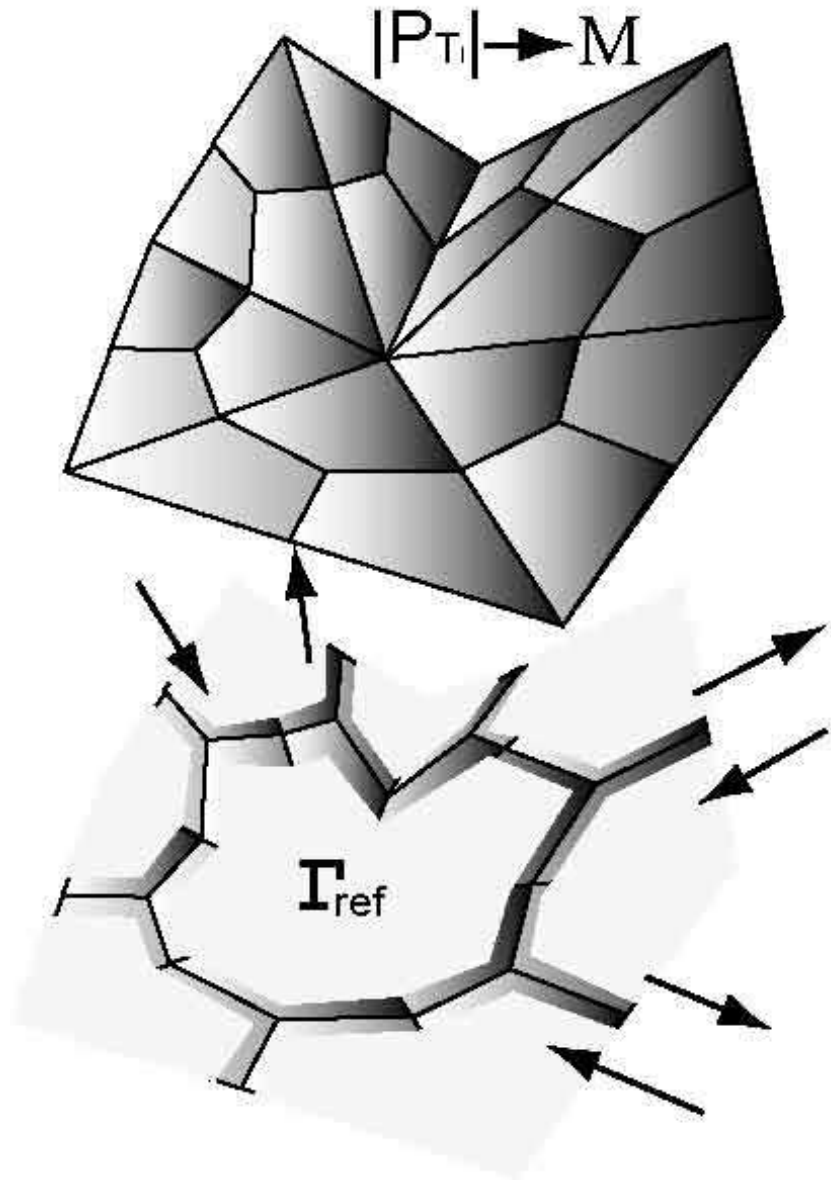}
\caption{The dual polytope around a vertex and it's edge refinement.}\label{fig7}
\end{center}
\end{figure}

\newpage

\begin{figure}[ht]
\begin{center}
\includegraphics[bb= 0 0 450 820,scale=.6]{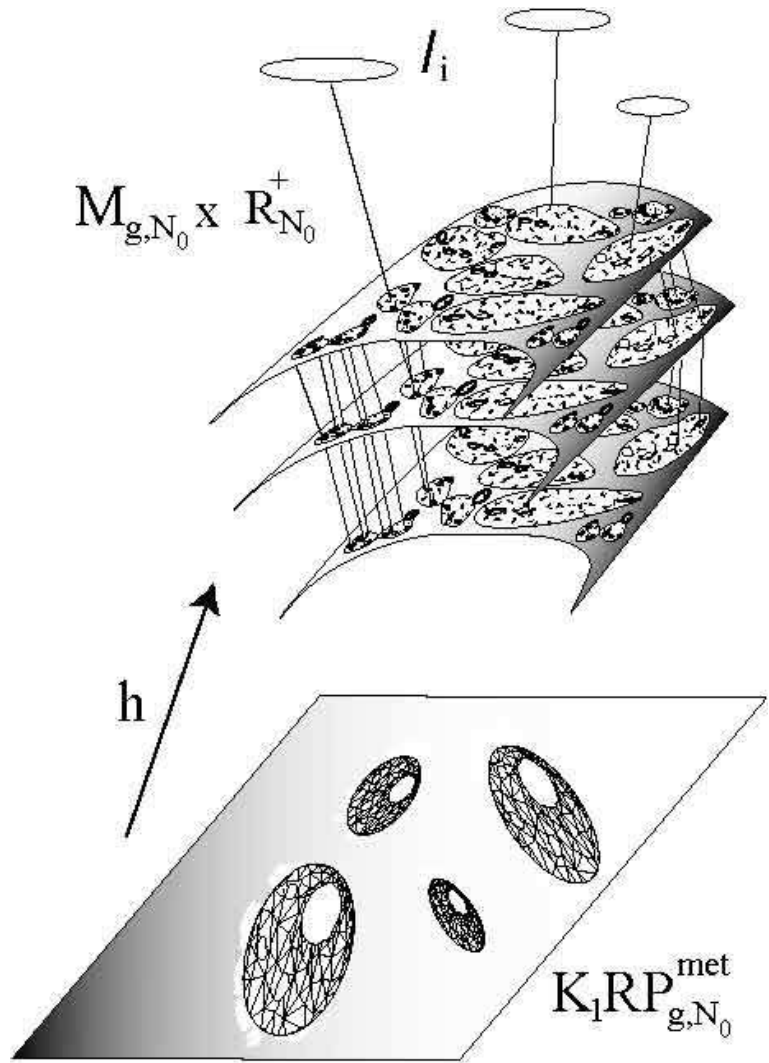}
\caption{The map $h$ associates to each ribbon graph an element of the decorated moduli space $\mathfrak{M}_{g,N_{0}}\times\Re^{+}_{N_{0}}$.}\label{fig8}
\end{center}
\end{figure}

\newpage

\begin{figure}[ht]
\begin{center}
\includegraphics[bb= 0 0 600 640,scale=.5]{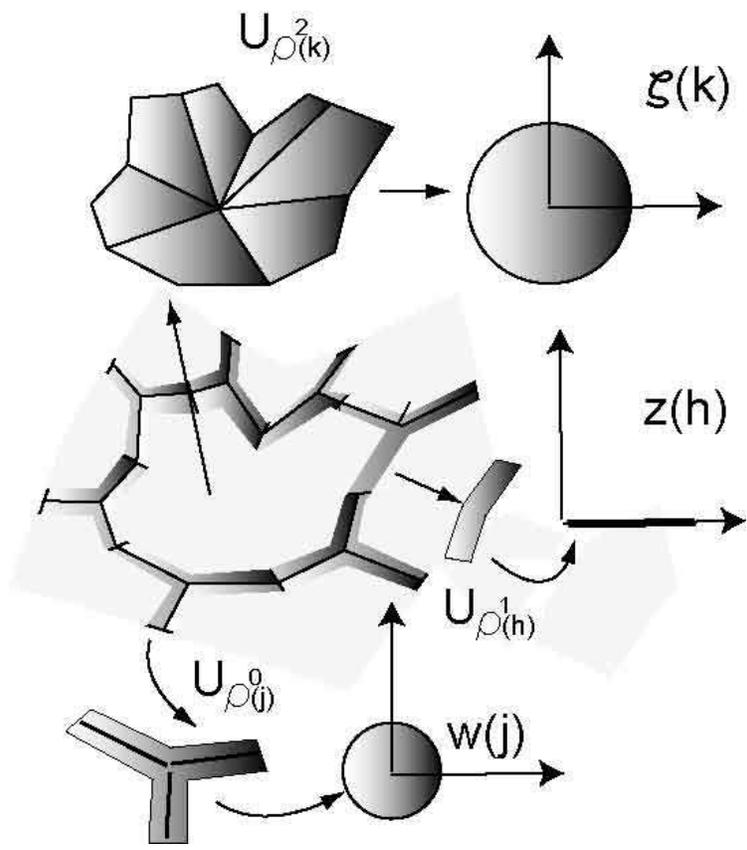}
\caption{The local presentation of uniformizing coordinates.}\label{fig9}
\end{center}
\end{figure}

\newpage

\begin{figure}[ht]
\begin{center}
\includegraphics[bb= 0 0 550 660,scale=.6]{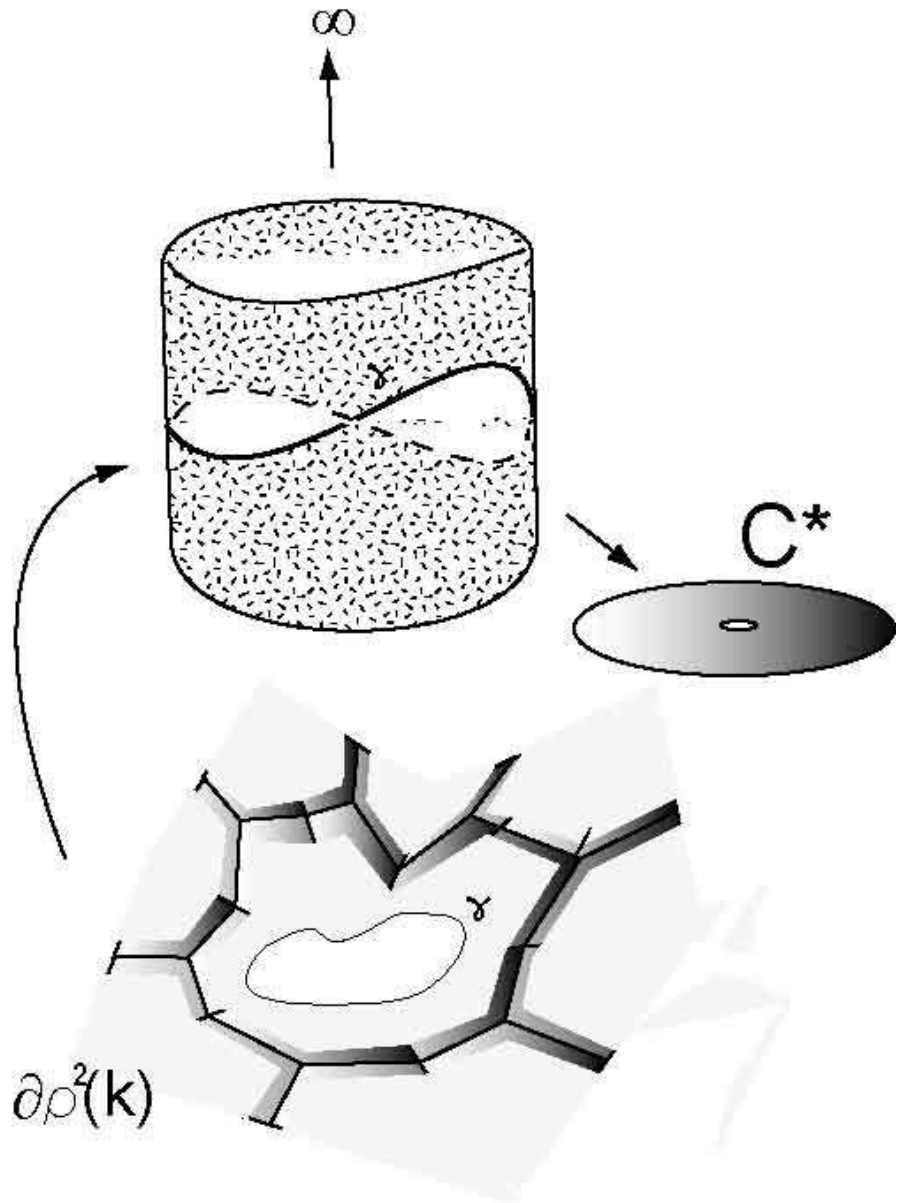}
\caption{The cylindrical metric associated with a ribbon graph.}\label{fig11}
\end{center}
\end{figure}

\newpage

\begin{figure}[ht]
\begin{center}
\includegraphics[bb= 0 0 555 590,scale=.6]{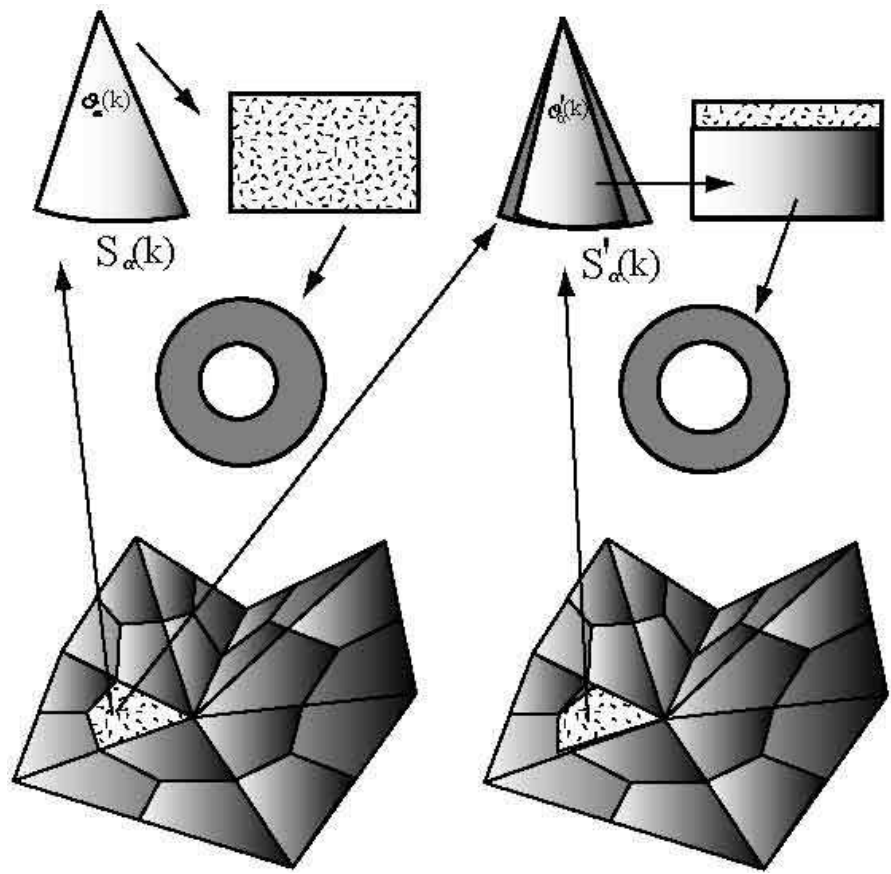}
\caption{The modular deformation of a conical sector in a Regge
polytope.}\label{fig12}
\end{center}
\end{figure}

\end{document}